\documentclass[12pt]{article}
\usepackage{amsfonts,amsmath,amssymb,epsf}
\usepackage{graphicx,color}
\newcommand{\be}{\begin{equation}}
\newcommand{\ee}{\end{equation}}

\newcommand{\ket}[1]{\left| #1 \right>}

\newcommand{\ba}{\begin{eqnarray}}
\newcommand{\ea}{\end{eqnarray}}

\newcommand{\mcl}{\mathcal}

\newcommand{\f}{\frac}

\newcommand{\no}{\nonumber \\ }

\newcommand{\ep}{\epsilon}

 \def\tr{\text{tr}}

\begin{document}

\begin{titlepage}
\thispagestyle{empty}

\begin{flushright}
YITP-15-54
\end{flushright}


\begin{center}
\noindent{\large \textbf{Quantum Entanglement of Fermionic Local Operators }}\\
\vspace{2cm}

Masahiro Nozaki $^{a}$,
Tokiro Numasawa $^{a}$ and
Shunji Matsuura $^{a}$
\vspace{1cm}

{\it
 $^{a}$Yukawa Institute for Theoretical Physics,
Kyoto University, Kyoto 606-8502, Japan\\}

\vskip 2em
\end{center}

\begin{abstract}
In this paper we study the time evolution of (Renyi) entanglement entropies for locally excited states in four dimensional free massless fermionic field theory. Locally excited states are defined by being acted by various local operators on the ground state. Their excesses are defined by subtracting (Renyi) entanglement entropy for the ground state from those for locally excited states. They finally approach some constant if the subsystem is given by half of the total space. They have spin dependence. They can be interpreted in terms of quasi-particles. 
\end{abstract}
\end{titlepage}
\newpage
\section{Introduction}

Quantum entanglement is an essential idea that distinguishes quantum physics from classical physics.
There has been a lot of work done to investigate various subfields of physics 
from a perspective of quantum entanglement.
For instance, 
the universal properties of quantum entanglement are used to characterize conformal field theories
\cite{Holzhey,Vidal,cc-04,Cal-Lev},
as well as topologically ordered phases \cite{Kitaev:2005dm,LW}.
The quantum entanglement also plays an essential role to understand the Hilbert space of gravity \cite{RT,VanRaamsdonk:2010pw}. 
While the most commonly used quantities for characterizing states
are correlation functions of operators,
 quantum entanglement focuses more directly on the Hilbert space.


Entanglement entropy is defined as the von Neumann entropy of a reduced density matrix $\rho_A$ for a subsystem $A$.
It has been observed that the dominant contribution of entanglement entropy in ground states of gapped systems comes from near boundary region of $A$.
This leads to the area law of entanglement entropy \cite{arealaw}.
For ground states of gapless systems, the contribution of the entanglement from a long distance does not necessarily become subdominant.
Indeed, in 2d conformal field theories (CFTs), entanglement entropy behaves universally as ${c\over 3} \log l$ where $c$ is the central charge and $l$ is the size of the subsystem $A$, suggesting that entanglement of any distance equally contributes to the entropy.
Recently some entanglement measures are introduced to describe the 
more detail structure of the entanglement entropy
\cite{econtour, Nozaki:2013wia}.

The above story is only for static (ground) states and much less is understood for
the dynamical aspects of the entanglement entropy as well as the entanglement entropy for excited states.
There are interesting questions about the dynamics of quantum entanglement, such as
how the quantum entanglement is created and 
propagates, how they distributes in space time.
Those are essential for understanding far-equilibrium states and their thermalization \cite{out of eq rev},
how efficiently quantum dynamics can be simulated in classical computers,
and where and how the black hole information goes.

In this paper, we study the time evolution of (R$\acute{e}$nyi) entanglement entropies 
by being acted by a fermionic local operator on the ground state.
First we explain quantum quenches which are 
a useful protocol to investigate the dynamical aspects of the entanglement. 
Suppose a system is prepared for a Hamiltonian $H(\lambda_0)$ which has an experimentally controllable parameter $\lambda_0$.
The state prepared for this system is given by a ground state.
Then at a certain time the parameter is shifted from $\lambda_0$ to $\lambda_1$. The prepared state is no longer a ground state of $H(\lambda_1)$
and the state undergoes a time evolution.
This kind of control is indeed possible in experiments such as cold atom systems \cite{entanglement-cold}.
When parameters are changed globally, these protocols are called as global quenches \cite{cc-global-quench}.
On the other hand, it is called as local quenches if parameters are changed locally \cite{cc-local-quench}.

Our protocol is similar to local quenches. In our setup, parameters of Hamiltonian are not changed. Instead of being changed them, a local operator $\mathcal{O}(t_0,x_0)$ acts on the ground state at certain time and creates entangled quasi-particle. In CFTs they propagate spherically at the speed of light.
If this local operator is inserted outside of $A$,
 a part of quasi-particles eventually enters into the subsystem $A$ and the rest part stays outside. 
 Since they are entangled, 
 it may create an entanglement between the subsystem $A$ and the rest of the system.
%
Indeed this kind of behavior, the increase of the entanglement and the saturation, has been 
 observed in various quantum field theories \cite{Alc,m1,m2,m3,He,He2,Cap2}.On the other hand, in holographic field theories the excesses of (R$\acute{e}$nyi) entanglement entropies do not saturate and keep to increase logarithmically even if time passes efficiently \cite{m3,Hart,Cap}.

The standard method for computing the entanglement entropy for ground states or thermal states in path integral is
the replica method: we compute the 
(R$\acute{e}$nyi) entanglement entropy $S_A^{(n)}=\frac{1}{1-n}\log \tr_A \rho^n_{A}$ first and then take $n\rightarrow 1$ limit, which gives 
entanglement entropy $S_A=-\tr_A \rho_{A}\log \rho_{A}$. Here $\rho_A=\tr_B \rho$ and we trace out the degrees of freedom outside $A$ (the region $B$). 

We consider free massless fermionic field theory in $4$ dimensional spacetime and take a half space as the subsystem $A$.
We generate an excitation  by being acted by local operator $\mathcal{O}$ at a distance $l$ away from the entangling surface and time $-t$.
The state prepared is given by a locally excited state. 
\begin{equation}
\ket{\Psi}=\mathcal{N}\mathcal{O}(-t,-l,{\bf x})\ket{0},
\end{equation}
where ${\bf x}=(x_2, x_3)$.
We define the excesses of (R$\acute{e}$nyi) entanglement entropies $\Delta S^{(n)}_A$ by subtracting (R$\acute{e}$nyi) entanglement entropies for the ground state from those for locally excited states.

They do not change for $t<l$, as expected from the causality.
The main difference is the final values of $\Delta S^{(n)}_A$: 
In the case of free massless fermionic field case, $\Delta S^{(n)}_A$ have spin dependence. Their density matrices can depend on the direction of spin because the probability with which (anti-)particles are included in $A$ can depend on the direction of spin. In the free massless scalar field theory, of course $\Delta S^{(n)}_A$ do not have such a dependence.

This paper is organized as follows.

In Sec.2, we explain our setup. 
In Sec.3, we explain the replica method and the analytic continuation which we perform.
In Sec.4 we derive the Green function in $4d$ free massless fermionic field theory on the replica space.
In Sec.5 we compute the time evolution of $\Delta S^{(n)}_A$.
In Sec.6 we explain the same thing in terms of quasi-particles.
In Sec.7, we conclude and discuss our results.

\section{Setup}
We study the time evolution of excesses of (R$\acute{e}$nyi) entanglement entropies for locally excited states which are defined by being acted by various local operators on the ground states in following setup. We consider the $4$ dimensional free massless fermionic theory,
\begin{equation}
S_{\text{fermion}} =- \int d^4x \bar{\psi}\gamma^{\mu}\partial_{\mu}\psi,
\end{equation}
where $\gamma^{\mu}=\{\gamma^t, \gamma^1, \gamma^2, \gamma^3 \}$ and $\bar{\psi}=i  \psi^{\dagger}\gamma^t$.

A local operator $\mathcal{O}$ acts on the ground state as in Figure.1. A locally excited state is given by 
\begin{equation}
\left|\Psi\right\rangle =\mathcal{N}\mathcal{O}(-t,-l,{\bf x})\left|0\right\rangle,
\end{equation}
where ${\bf x}=(x^2, x^3)$ and $\mathcal{N}$ is a normalization constant. 

The subsystem $A$ is given by a half of the total space, $x^1\ge 0$ as in Figure.1. We trace out the degrees of freedom in the complement space $B$ outside $A$ and define a reduced density matrix,
\begin{equation}
\rho^{EX}_A =\tr_B\rho^{EX}
\end{equation} 
where $\rho=\left|\Psi\right\rangle \left\langle \Psi \right|$.

By using this reduced density matrix, (R$\acute{e}$nyi) entanglement entropy for locally excited state is defined by
\begin{equation}
S^{(n) EX}_A=\frac{1}{1-n}\log{\left[\tr_A\left(\rho_A^{ EX}\right)^n\right]}. 
\end{equation}
By using the reduced density matrix for the ground state $\rho_A^G=tr_B\left|0\right\rangle \left\langle 0 \right|$, (R$\acute{e}$nyi) entanglement entropy for the ground state is defined by
\begin{equation}
S^{(n) G}_A=\frac{1}{1-n}\log{\left[\tr_A\left(\rho_A^G\right)^n\right]}. 
\end{equation}
We define the excess of (R$\acute{e}$nyi) entanglement entropy by subtracting $S^{(n) G}_A$ from $S^{(n) EX}_A$,
\begin{equation}
\Delta S_A^{(n)} = S^{(n) EX}_A-S^{(n) G}_A.
\end{equation}

We will study the time evolution of $\Delta S^{(n)}_A$ in following sections.
\begin{figure}
  \centering
  \includegraphics[width=8cm]{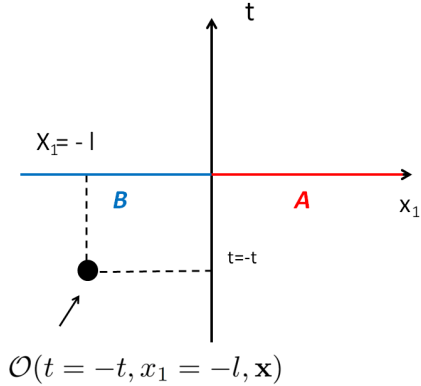}
  \caption{The location of an local operator and the subsystem $A$ in the Minkowski spacetime.}
\end{figure}
\section{Replica Trick}
\subsection{Locally Excited States}
In this section, we explain the replica method for locally excited states.
The states we are considering are given as follows:
\be
\left|\Psi\right\rangle = \mcl{N} e^{-iHt }e^{-\ep H} \mcl{O}(-l,{\bf x} )\left|0\right\rangle.
\ee
Here $\mcl{O}(-l,{\bf x})$ is a local operator in Shr\"{o}dinger picture and we introduce the regularization factor $e^{-\ep H}$ 
because the norm of the state $\left|\Psi\right\rangle $ is divergent  without $e^{-\ep H}$ and $\left|\Psi\right\rangle $ does not belong
to the Hilbert space. This corresponds to smearing the point like excitation.

The density matrix $\rho$ is given by 
\ba
\rho &=& \mcl{N}^{2} e^{-iHt} e^{-\ep H} \mcl{O}(-l,{\bf x}) \left|0\right\rangle \left\langle0\right|\mcl{O} (-l,{\bf x})^{\dagger} e^{- \ep H}
e^{iHt} \no
&=& \mcl{N}^{2} \mcl{O}(\tau_e,-l,{\bf x})\left|0\right\rangle \left\langle0\right| \mcl{O}^{\dagger} (\tau_l,-l,{\bf x}) 
\ea 
Here $\mcl{O}(\tau,-l,{\bf x})$ is a local operator in Heisenberg picture and in the second line we introduce complex times $\tau_{e} =-\ep -it$  and $\tau_l = \ep -it$.
In the calculation of (R\'enyi) entanglement entropy, we first treat complex times $\tau_e$ and $\tau_l$ as if they are real parameters and finally we analytically continue to complex values.

\subsection{Replica method}

Now we explain the replica method for locally excited states. In the path integral formalism, we can express the wave functional for locally excited states is given by 
\be
\Psi(\phi(x)) = \left(Z_1^{EX}\right)^{-\frac{1}{2}}\left\langle\phi| \Psi \right\rangle = \int _{\phi(\tau = - \infty,x^i ) } ^{\phi(\tau = 0,x^i) = \phi(x^i)} \mcl{D}\phi \ \mcl{O}(\tau_e,-l,{\bf x}) e^{- S[\phi]}.
\ee  
In the same manner, the we can express the bra vector as follows:
\be
\Psi^*(\phi(x)) = \left(Z_1^{EX}\right)^{-\frac{1}{2}}\left\langle\Psi| \phi \right\rangle = \int ^{\phi(\tau =  \infty,x^i ) } _{\phi(\tau = 0,x^i) = \phi(x^i)} \mcl{D} \phi \ \mcl{O}^{\dagger}(\tau_l,-l,{\bf x}) e^{- S[\phi]}.
\ee
Then, the density matrix for total system is given by the path integral on the space which has boundary at $\tau=+0$ and $\tau=-0$:
\begin{equation}
\begin{split}
[\rho_{tot}]_{\phi \phi'} &= \left\langle\phi|\Psi\right\rangle\left\langle\Psi|\phi'\right\rangle \\ &=(Z_1^{EX})^{-1} \int_{\phi(\tau=+0,x^i)=\phi'(x^i),\phi(\tau=-0,x^i)=\phi(x^i)} \mcl{D} \phi \ \mcl{O}^{\dagger}(\tau_l,-l,{\bf x})  \mcl{O}(\tau_e,-l,{\bf x}) e^{- S[\phi]},
\end{split}
\end{equation}
where $Z_1^{EX}$ appears in order to keep $\tr \rho=1$  and it is given by
\be
Z_1^{EX} = \int ^{\phi(\tau=\infty,x^i)}_{\phi(\tau=-\infty,x^i)} \mcl{D} \phi \mcl{O}^{\dagger}(\tau_l,-l,{\bf x})   \mcl{O}(\tau_e,-l,{\bf x}) e^{- S[\phi]}.
\ee
Partial trace corresponds to sawing the region which was traced out, so the reduced density matrix is given by
\begin{equation}
\begin{split}
[\rho_A^{EX}]_{\phi \phi'} = (Z_1^{EX})^{-1} \int ^{\phi(\tau=\infty,x^i)}_{\phi(\tau=-\infty,x^i)} &\mcl{D} \phi  \mcl{O}^{\dagger}(\tau_l,-l,{\bf x}) \mcl{O}(\tau_e,-l,{\bf x}) e^{- S[\phi]} \\
&\times
\prod_{x\in A} \delta(\phi(+0,x^i)-\phi'(x^i))\cdot\delta(\phi(-0,x^i)-\phi(x^i)).
\end{split}
\end{equation}
From this we can see that the only difference between the reduced density matrix for ground states and that for locally excited states is the insertion
of local operators $\mcl{O}(\tau_e)$ and $\mcl{O}(\tau_l)$.
We need to insert two local operators in each sheet, so finally we need to insert $2n$
local operators in the $n$-sheeted manifold $\Sigma_n$ which is constructed of $n$ flat spaces and it has a conical singularity on the entangling surface as in Figure.2.

Then, the $\tr (\rho^{EX}_A)^n$ is given by the partition function with the insertion of $2n$ local operators:
\be
\tr (\rho^{EX}_A)^n = (Z_1^{EX})^{-n} \int \mcl{D} \phi   \mcl{O}^{\dagger}(r_l,\theta_l^n) \mcl{O}(r_e,\theta_e^n)  \cdots \mcl{O}^{\dagger}(r_l,\theta_l^1) \mcl{O}(r_e,\theta_e^1)e^{- S[\phi]}. \label{expf2}
\ee
where we introduce the polar coordinate $(r,\theta)$ on $(\tau,x_1)$ plane , the region of $\theta$ is given by $0<\theta < 2\pi n$ and $\theta_{e,l} ^k  = \theta_{e,l} ^1  + 2 \pi (k-1) $, see Figure.2.
 \begin{figure}
 \begin{center}
  \includegraphics[width=100mm]{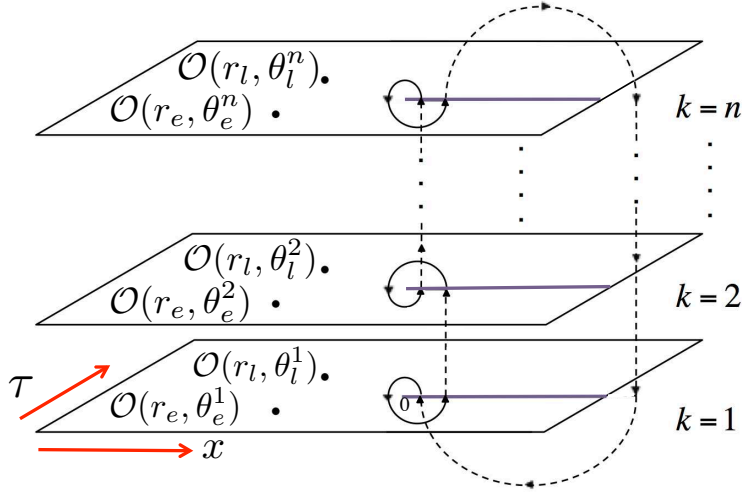}
 \end{center}
 \caption{$n$-sheeted manifold with operator insertion}
 \label{branched}
\end{figure}
 (\ref{expf2}) is almost the correlation function and the only difference is that 
right hand side is divided by $(Z_1^{EX})^n$, not by $Z_n$ where $Z_n$ is the partition function on $n$-sheeted manifold $\Sigma_n$ without any operator insertion. 
If we consider the difference between R\'enyi entanglement entropy for excited states and that for the ground state, we can find that it is expressed by the 
correlation function on $n$-sheeted manifold:
\begin{equation}
\begin{split}
&\Delta S_A^{(n)} = \f{1}{1-n} \Big( \log \f{\tr_A(\rho_A^{EX})^n}{(\tr_A \rho_A^{EX})^n} - \log \f{\tr_A(\rho_A^{G})^n}{(\tr_A \rho_A^{G}){}^n} \Big) = \f{1}{1-n} \Big( \log \f{Z_n^{EX}}{Z_n} - n\log \f{Z_1^{EX}}{Z_1^{G}} \Big) \\
&= \f{1}{1-n} (\log \left\langle  \mcl{O}^{\dagger}(r_l,\theta_l^n) \mcl{O}(r_e,\theta_e^n)  \cdots \mcl{O}^{\dagger}(r_l,\theta_l^1) \mcl{O}(r_e,\theta_e^1) \right\rangle_{\Sigma_n} - n \log \left\langle\mcl{O}(r_e,\theta_e^1) \mcl{O}^{\dagger}(r_l,\theta_l^1) \right\rangle_{\Sigma_1} ).
\end{split}
\end{equation}
In this way, we can express the difference of (R\'enyi) entanglement entropy using 
the correlation function on $n$-sheeted manifold.

\section{Propagator}
In conformal field theories, (R$\acute{e}$nyi) entanglement entropy for a ground state is invariant under the conformal transformation. In order to preserve its conformal symmetry in the replica trick, the action on $\Sigma_n$ is given by 
\begin{equation}\label{cvc}
 S=\int_{\Sigma_n} d V \bar{\psi} \Gamma^{\mu} \nabla_{\mu}\psi,
\end{equation}
where $\Gamma^i (i=0,1,2,3)$ obeys the clifford algebra ($\left\{\Gamma^i, \Gamma^j\right\}=2 \delta_{i,j}{\bf 1}$) and ${\bf 1}$ is the identity.

In this case, $\Sigma_n$ is given by the flat space except for the origin. We introduce a polar coordinate and this geometry is described by 
\begin{equation}
ds^2 =dr^2+ r^2 d\theta^2+ dx_2^2 +dx_3^2.
\end{equation}

The two point function of $\psi$ and $\bar{\psi}$ is defined by 
\begin{equation}
-\left\langle \mathcal{T}\psi_a(r, \theta, {\bf x})\bar{\psi}_b(r_2, \theta_2, {\bf x}_2)\right \rangle =S_{ab}(r, r_2, \theta, \theta_2, {\bf x}, {\bf x}_2).
\end{equation}

$S_{ab}$ obeys the equation of motion as follows,
\begin{equation}
\left( e^{\mu}_i \Gamma^i \partial_{\mu}+\frac{\Gamma^0}{2r}\right)S(r, r_2, \theta, \theta_2, {\bf x}, {\bf x}_2)=- {\bf 1}\frac{\delta\left(r-r2\right)\delta\left(\theta-\theta_2\right)\delta\left({\bf x}-{\bf x_2}\right)}{r}.
\end{equation}

$S_{ab}$ can be rewritten by 
\begin{equation}\label{s}
S_{ab}=\left(e^{\mu}_i \Gamma^i \partial_{\mu}+\frac{\Gamma^0}{2r}\right)g(r, r_2, \theta, \theta_2, {\bf x}, {\bf x}_2).
\end{equation}

$g$ is defined by
\begin{equation}\label{g}
g(x,x')={\bf 1}Re H(r, r_2, \theta, \theta_2, {\bf x}, {\bf x}_2)+\Gamma^0 \Gamma^1 Im H (r, r_2, \theta, \theta_2, {\bf x}, {\bf x}_2).
\end{equation}
where $H(x,x')$ is given by 
\begin{equation}
H(r, r_2, \theta, \theta_2, {\bf x}, {\bf x}_2)=\exp{\left(i\frac{\theta-\theta'}{2}\right)}G(r, r_2, \theta, \theta_2, {\bf x}, {\bf x}_2).
\end{equation}

$G(r, r_2)$ obeys the following equation of motion,
\begin{equation}
\begin{split}
\left(\frac{\partial^2}{\partial r^2}+\frac{1}{r}\frac{\partial}{\partial r}+\frac{1}{r^2}\frac{\partial^2}{\partial \theta^2}+\frac{\partial^2}{(\partial x^2)^2}+\frac{\partial^2}{(\partial x^3)^2}\right)&G(r, r_2, \theta, \theta_2, {\bf x}, {\bf x}_2)= \\
&-\frac{\delta\left(r-r_2\right)\delta\left(\theta-\theta_2\right)\delta\left({\bf x}-{\bf x_2}\right)}{r}.
\end{split}
\end{equation}

Next we consider the boundary condition for $\psi(x)$\footnote{By using the map in \cite{chm}, entanglement entropy for the spherical subsystem is map to the thermal entropy in the hyperbolic space. This boundary condition in (\ref{bc}) agrees with the one which is imposed on fermionic field along $S^1$ direction.}.
It is given by,
\begin{equation}\label{bc}
\psi(r, \theta+2n \pi, {\bf x})= -\psi (r, \theta, {\bf x}).
\end{equation}
Then the boundary condition for two point function of $\psi$ is given by 
\begin{equation}
S_{ab}(r,  \theta+2n \pi,  {\bf x})=- S_{ab}(r,  \theta,  {\bf x}).
\end{equation}

The boundary condition for the green function $G$ is given by
\begin{equation} \label{bcg}
G(r, \theta+ 2 n \pi, {\bf x})=e^{i (-n+1)\pi} G(r, \theta , {\bf x}).
\end{equation}

\subsection{Computation of Proapgator}

Let's compute  the propagator $G(r, r', \theta, \theta')$.
It can be expanded by the eigenfunctions $V(r, \theta, {\bf x})$ which are defined by
\begin{equation}
\mathcal{L}V(r, \theta, {\bf x})=-(k^2+{\bf k}^2) V(r, \theta, {\bf x}),
\end{equation}  
where $\mathcal{L}= \left(\frac{\partial^2}{\partial r^2}+\frac{1}{r}\frac{\partial}{\partial r}+\frac{1}{r^2}\frac{\partial^2}{\partial \theta^2}+\frac{\partial^2}{ (\partial x^2)^2}+\frac{\partial^2}{(\partial x^3)^2}\right)$.

After expanding it, it is given by 
\begin{equation}
\begin{split}
G(x, x')&=\frac{1}{2n \pi}\frac{1}{(2\pi)^2}\sum^{\infty}_{l=-\infty} \int d {\bf k} \int^{\infty}_0 dk k \frac{J_{\left|\frac{-n+1+2l}{2n}\right|}(k r) J_{\left|\frac{-n+1+2l}{2n}\right|}(k r')}{k^2+{\bf k}^2} \\
&~~~~~~~~\times e^{i\left(\frac{-n+1+2l}{2n}\right)(\theta- \theta')}e^{i{\bf k}({\bf x}-{\bf x}')},
\end{split}
\end{equation}
where $J_{\nu}(x) $ is the Bessel function of the first kind.
We can rewrite this green function as in \cite{m1,m2,m3,Dow,sac2,Line}. It is given by
\begin{equation}
G(x, x')=\frac{1}{8\pi^2n r r'} \sum_{l=-\infty}^{\infty}\frac{e^{-\left|\frac{-n+1+2l}{2n}\right|t_0 +i\left(\frac{-n+1+2l}{2n}\right)(\theta- \theta')}}{\sinh{t_0}},
\end{equation}
where $t_0$ is defined by
\begin{equation}
\cosh{t_0}=\frac{r^2+r'^2+\left|{\bf x}-{\bf x}' \right|^2}{2 r r'}.
\end{equation}

If $n$ is odd, the green function is given by
\begin{equation}\label{odd}
G(x, x') =\frac{1}{8\pi^2 n r r'} \frac{\sinh{\frac{t_0}{n}}}{\sinh{t_0}\left(\cosh{\left(\frac{t_0}{n}\right)}-\cos{\left(\frac{\theta-\theta'}{n}\right)}\right)}
\end{equation}
In this case, the boundary condition (\ref{bcg}) is given by 
\begin{equation}
G(r, \theta +2n \pi, {\bf x})= G(r, \theta, {\bf x}).
\end{equation}
The green function given by (\ref{odd}) obeys this boundary condition.

If $n$ is even, the green function is 
\begin{equation}\label{even}
G(x, x')=\frac{1}{4\pi^2 n r r'}\frac{\cos{\left(\frac{\theta-\theta'}{2n}\right)}\sinh{\frac{t_0}{2n}}}{\sinh{t_0}\left(\cosh{\left(\frac{t_0}{n}\right)}-\cos{\left(\frac{\theta-\theta'}{n}\right)}\right)}.
\end{equation}
In this case, boundary condition given by (\ref{bcg}) is given by 
\begin{equation}
G(r, \theta+2n \pi, {\bf x})=- G(r, \theta, {\bf x}).
\end{equation}

\subsection{Analytic Continuation to Real Time}
Up to here, we consider propagators in the Euclidean space. Two local operators are located as in Figure.3. We would like to study the time evolution of (R$\acute{e}$nyi) entanglement entropy. Therefore we perform an analytic continuation to real time as follows,
\begin{equation}\label{ac}
\begin{split}
&\tau_l = \epsilon -i t, \\
&\tau_e=-\epsilon -i t,
\end{split}
\end{equation}
where in lorentzian spacetime $\epsilon$ is a cutoff parameter which regulates the divergence when a local operator contacts with another.

After performing it, parameters in Euclidean space are related to those in lorentzian spacetime as follows,
\begin{equation}
\begin{split}
&r^2 = l^2-t^2+\epsilon^2-2i\epsilon t, \\
&r_2^2 = l^2-t^2+\epsilon^2+2i\epsilon t, \\
&\cos{\left(\theta-\theta_2\right)}=\frac{l^2-\epsilon^2-t^2}{V^2}, \\
&\sin{\left(\theta-\theta_2\right)}=-\frac{2\epsilon t}{V^2}, \\
&V^2=\sqrt{(l^2-t^2+\epsilon^2)^2+4\epsilon^2 t^2}. \\
\end{split}
\end{equation}

\begin{figure}
  \centering
  \includegraphics[width=8cm]{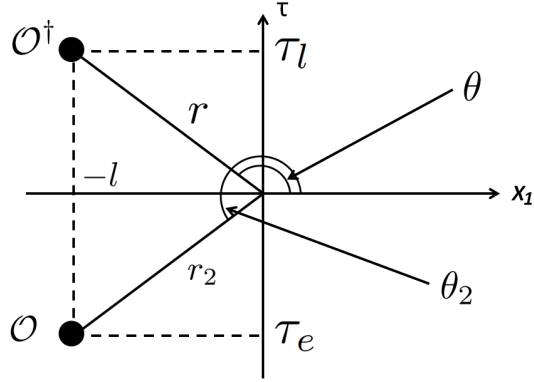}
  \caption{The location of operators in Euclidean space.}
\end{figure}
\subsection{Dominant Propagators}
After performing the analytic continuation in (\ref{ac}), we take the limit $\epsilon \rightarrow 0$. A few propagators dominantly contribute to $n$ point functions. We call them dominant propagators. They are $\mathcal{O}(\epsilon^{-3})$. If $t\le l$, dominant propagators on $\Sigma_1$ are given by 
\begin{equation}
\begin{split}
&S_{ab}(r,r_2,\theta- \theta_2)=\frac{1}{16\pi^2\sqrt{l^2-t^2}\epsilon^3}\left[it\Gamma^0_{ab}+l\Gamma^1_{ab}\right], \\
&S_{ab}(r,r_2,\theta_2- \theta)
=\frac{-1}{16\pi^2\sqrt{l^2-t^2}\epsilon^3}\left[it\Gamma^0_{ab}+l\Gamma^1_{ab}\right]. \\
\end{split}
\end{equation} 
On the other hand, If $t > l$, they are given by 
\begin{equation}
\begin{split}
&S_{ab}(r,r_2,\theta- \theta_2)=\frac{1}{16\pi^2\sqrt{t^2-l^2}\epsilon^3}\left[il\Gamma^0_{ab}+t\Gamma^1_{ab}\right], \\
&S_{ab}(r,r_2,\theta_2- \theta)
=\frac{-1}{16\pi^2\sqrt{t^2-l^2}\epsilon^3}\left[il\Gamma^0_{ab}+t\Gamma^1_{ab}\right], \\
\end{split}
\end{equation}

If $t \le l$, dominant propagators on $\Sigma_{n>1}$ are red arrows in Figure.4.
They are given by
\begin{equation}
\begin{split}
&S_{ab}(r, r_2, \theta- \theta_2)=\frac{1}{16\pi^2\sqrt{l^2-t^2}\epsilon^3}\left[it\Gamma^0_{ab}+l\Gamma^1_{ab}\right], \\
&S_{ab}(r_2, r, \theta_2- \theta)
=\frac{-1}{16\pi^2\sqrt{l^2-t^2}\epsilon^3}\left[it\Gamma^0_{ab}+l\Gamma^1_{ab}\right]. \\
\end{split}
\end{equation}
If $t>l$, they are red arrows and blue arrows in Figure.4.
They are given by 
\begin{equation}\label{p1}
\begin{split}
&S_{ab}(r, r_2 ,\theta- \theta_2)
=\frac{(t+l)^2}{64\pi^2t\sqrt{t^2-l^2}\epsilon^3}\left[i\Gamma^0_{ab}+\frac{2t-l}{t}\Gamma^1_{ab}\right], \\
&S_{ab}(r_2, r, \theta_2- \theta)
=\frac{-(t+l)^2}{64\pi^2t\sqrt{t^2-l^2}\epsilon^3}\left[i\Gamma^0_{ab}+\frac{2t-l}{t}\Gamma^1_{ab}\right], \\
&S_{ab}(r, r_2 ,\theta- \theta_2+2\pi)
=\frac{(t-l)^2}{64\pi^2t\sqrt{t^2-l^2}\epsilon^3}\left[i\Gamma^0_{ab}-\frac{2t+l}{t}\Gamma^1_{ab}\right], \\
&S_{ab}(r_2, r ,\theta_2- \theta-2\pi)
=\frac{-(t-l)^2}{64\pi^2t\sqrt{t^2-l^2}\epsilon^3}\left[i\Gamma^0_{ab}-\frac{2t+l}{t}\Gamma^1_{ab}\right], \\
&S_{ab}(r, r_2 ,\theta- \theta_2-2(n-1)\pi)
=\frac{-(t-l)^2}{64\pi^2t\sqrt{t^2-l^2}\epsilon^3}\left[i\Gamma^0_{ab}-\frac{2t+l}{t}\Gamma^1_{ab}\right], \\
&S_{ab}(r_2, r ,\theta_2- \theta+2(n-1)\pi)
=\frac{(t-l)^2}{64\pi^2t\sqrt{t^2-l^2}\epsilon^3}\left[i\Gamma^0_{ab}-\frac{2t+l}{t}\Gamma^1_{ab}\right]. \\
\end{split}
\end{equation}

\begin{figure}[h]
 \begin{minipage}{0.5\hsize}
  \begin{center}
   \includegraphics[width=70mm]{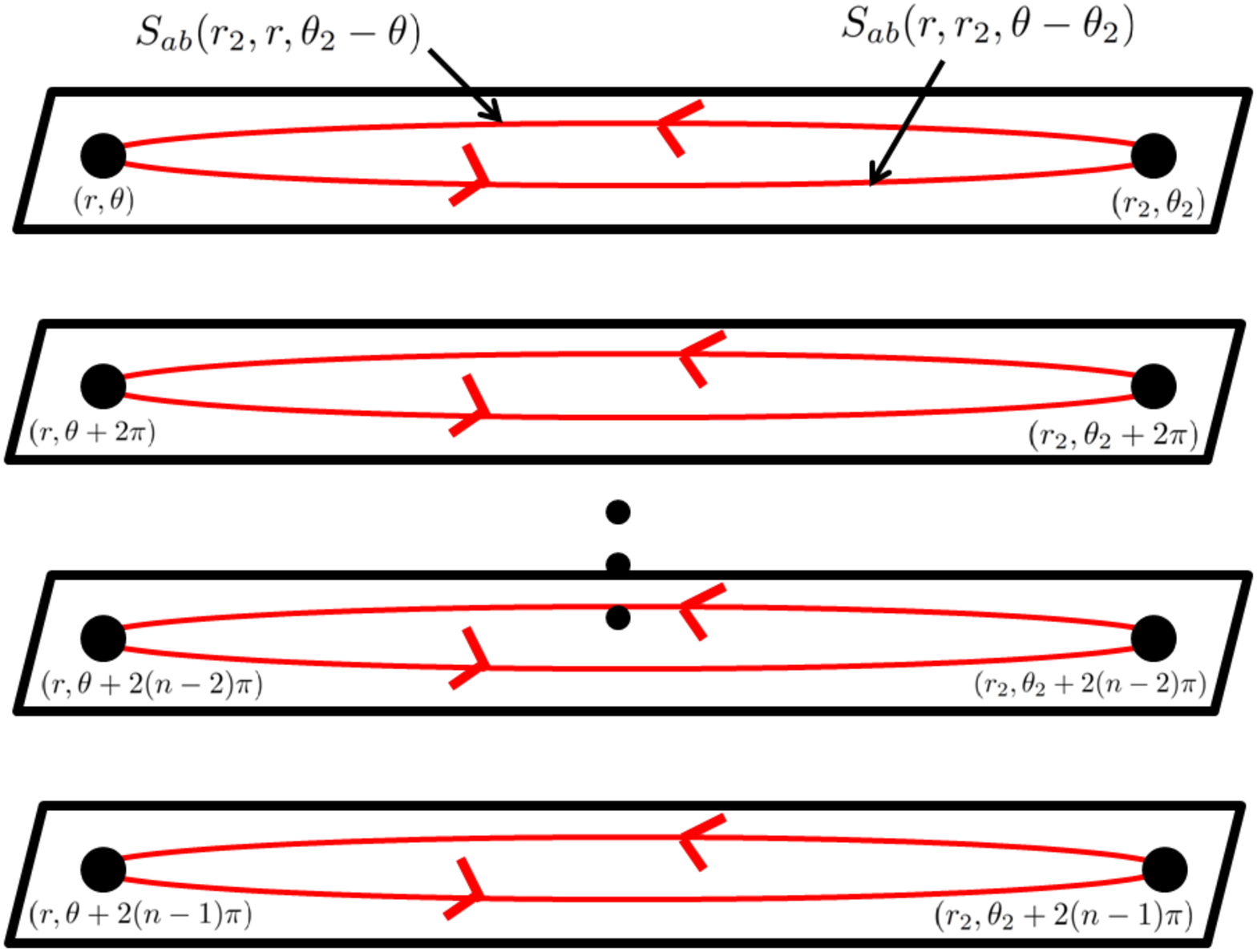}
  \end{center}
 \end{minipage}
  \begin{minipage}{0.30\hsize}
  \begin{center}
   \includegraphics[width=70mm]{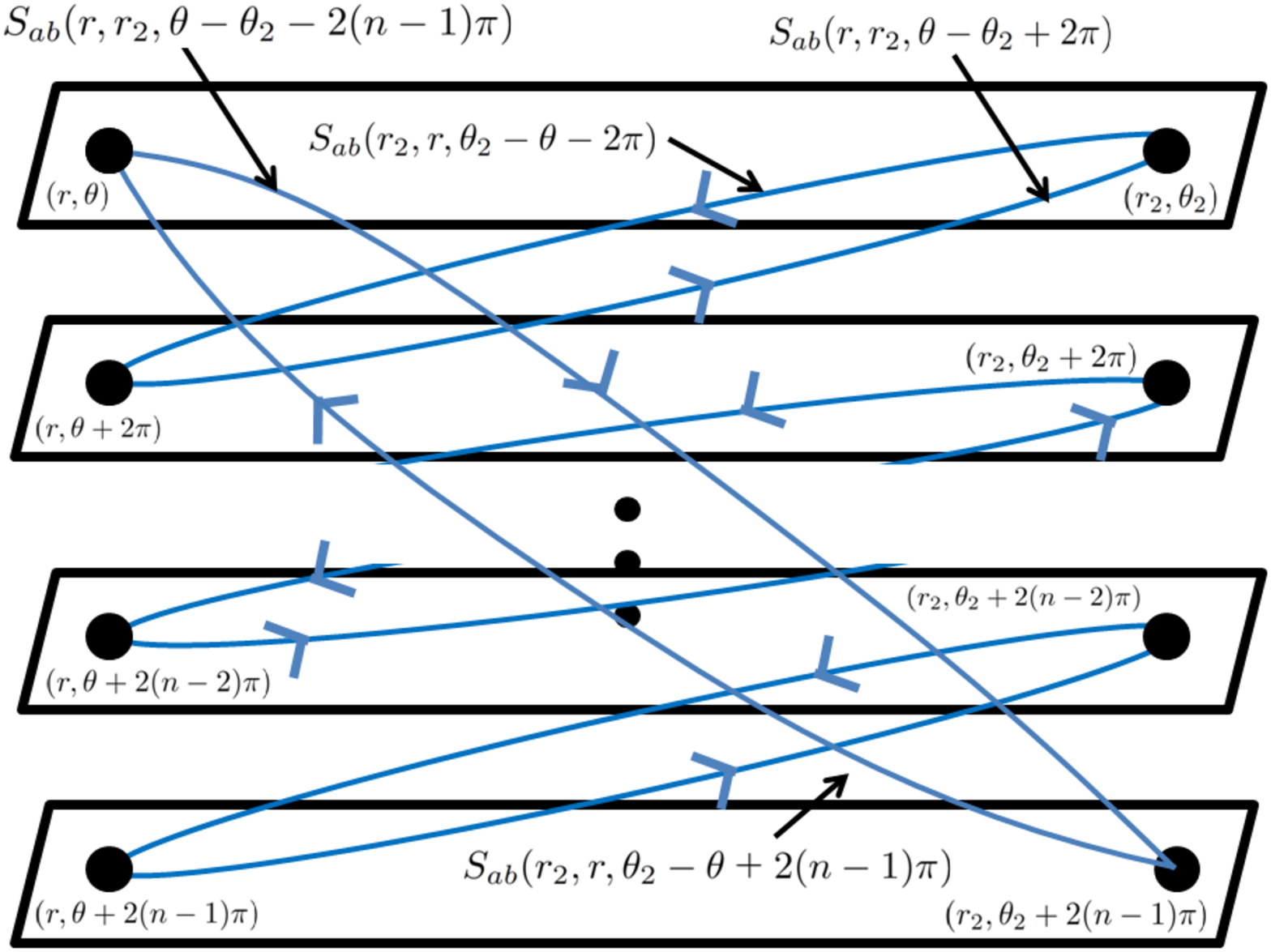}
  \end{center}
 \end{minipage}
 \caption{The schematic description of dominant propagators. $S_{ab}(r, r_2, \theta-\theta_2)$ and $S_{ab}(r_2, r, \theta_2-\theta)$ correspond to the right arrow and the left arrow respectively. }\label{ph}
\end{figure}
\subsubsection{Propagators in the Cartesian Coordinate }
Given locally excited states are defined by being acted by local operators on the ground state in Cartesian coordinate. Therefore we explain the relation between propagators in (\ref{cvc}) and them in the Cartesian coordinate.
In the Cartesian coordinate, the action in free massless fermionic theory is given by 
\begin{equation}
S=\int d^4x\bar{\psi}'\gamma^{\mu}\partial_{\mu}\psi',
\end{equation}
where $\gamma^{\mu}=\{\gamma^0, \gamma^1, \gamma^2, \gamma^3\}$.

When we introduce a polar coordinate $x^1=r \cos{\theta}, x^0=r \sin{\theta}$ and redefine the fermionic field by $\psi' (r,\theta)=e^{-\frac{\gamma^1\gamma^0}{2}\theta}\psi(r,\theta)$, it is rewritten by
\begin{equation}
S=\int dr \int d\theta r \int d^2{\bf x} \bar{\psi}\left[\Gamma^0\partial_r+\Gamma^1\frac{\partial_{\theta}}{r}+\frac{\Gamma^0}{2r}+\Gamma^{{\bf x}}\partial_{{\bf x}}\right]\psi,
\end{equation}
where $\Gamma^0=\gamma^1, \Gamma^1=\gamma^0, \Gamma^{{\bf x}}=\gamma^{{\bf x}}$.
\begin{equation}
\psi'(r, \theta)=\cos{\left(\frac{\theta}{2}\right)}\psi(r, \theta)+\sin{\left(\frac{\theta}{2}\right)}\gamma^0\gamma^1\psi(r, \theta).
\end{equation}
After perfprming this map, the boundary condition for $\psi'$\footnote{When $n=1$, $\psi'$ is regular at the origin.} is mapped to 
\begin{equation}
\psi'(r, \theta+2n \pi)=(-1)^{n+1}\psi'(r, \theta).
\end{equation}

We define propagators by
\begin{equation}
\begin{split}
&-\left\langle \mathcal{T} \psi'_a(r, \theta) \bar{\psi}_b'(r_2, \theta_2)\right \rangle = S'_{ab}(r, r_2, \theta-\theta_2), \\
&-\left\langle \mathcal{T} \psi'_a(r, \theta) \psi_b'^{\dagger} (r_2, \theta_2)\right \rangle = S'_{ac}(r, r_2, \theta-\theta_2)\gamma^0_{cb} = V_{ab}(r, r_2, \theta-\theta_2). \\
\end{split}
\end{equation}

After performing this map, we also perform the analytic continuation in (\ref{ac}) and take the limit $\epsilon \rightarrow 0$. 
After that, only a few propagators can contribute to $n$ point function dominantly. They are $\mathcal{O}(\epsilon^{-3})$.
In any time, dominant propagators on $\Sigma_1$ are given by 
\begin{equation}\label{dpe}
\begin{split}
S'_{ab}(r, r_2, \theta-\theta_2)=\frac{-i}{16\pi^2\epsilon^3}\gamma^{t}_{ab}, \\
S'_{ab}(r_2, r, \theta_2-\theta)=\frac{i}{16\pi^2\epsilon^3}\gamma^{t}_{ab}, \\
V_{ab}(r, r_2, \theta-\theta_2)=\frac{-1}{16\pi^2\epsilon^3}{\bf 1}_{ab}, \\
V_{ab}(r_2, r, \theta_2-\theta)=\frac{1}{16\pi^2\epsilon^3}{\bf 1}_{ab}, \\
\end{split}
\end{equation}
where ${\bf 1}$ is identity. $\gamma^0$ is changed to $i \gamma^t \left((\gamma^t)^2=-{\bf 1}\right)$. 

On the other hand, if $t \le l$ those on $\Sigma_{n>1}$ are the same as in (\ref{dpe}).
In the region $t>l$, they are given by 
\begin{equation}\label{p2}
\begin{split}
&S'_{ab}(r, r_2, \theta-\theta_2)=\frac{-i(t+l)}{64\pi^2t\epsilon^3}\left[\frac{t-l}{t}\gamma^1_{ab}+2\gamma^t_{ab}\right], \\
&S'_{ab}(r_2, r, \theta_2-\theta)=\frac{i(t+l)}{64\pi^2t\epsilon^3}\left[\frac{t-l}{t}\gamma^1_{ab}+2\gamma^t_{ab}\right], \\
&S'_{ab}(r, r_2, \theta-\theta_2+2\pi)=\frac{i(t-l)}{64\pi^2t\epsilon^3}\left[\frac{t+l}{t}\gamma^1_{ab}-2\gamma^t_{ab}\right], \\
&S'_{ab}(r_2, r, \theta_2-\theta-2\pi)=-\frac{i(t-l)}{64\pi^2t\epsilon^3}\left[\frac{t+l}{t}\gamma^1_{ab}-2\gamma^t_{ab}\right], \\
&S'_{ab}(r, r_2, \theta-\theta_2-2(n-1)\pi)=(-1)^{n-1}\frac{i(t-l)}{64\pi^2t\epsilon^3}\left[\frac{t+l}{t}\gamma^1_{ab}-2\gamma^t_{ab}\right], \\
&S'_{ab}(r_2, r, \theta_2-\theta+2(n-1)\pi)= (-1)^{n} \frac{i(t-l)}{64\pi^2t\epsilon^3}\left[\frac{t+l}{t}\gamma^1_{ab}-2\gamma^t_{ab}\right], \\
\end{split}
\end{equation}
and
\begin{equation}\label{p3}
\begin{split}
&V_{ab}(r, r_2, \theta-\theta_2)=-\frac{(t+l)}{64\pi^2t\epsilon^3}\left[-\frac{t-l}{t}\left(\gamma^1\gamma^t\right)_{ab}+2{\bf 1}_{ab}\right], \\
&V_{ab}(r_2, r, \theta_2-\theta)=\frac{(t+l)}{64\pi^2t\epsilon^3}\left[-\frac{t-l}{t}\left(\gamma^1\gamma^t\right)_{ab}+2{\bf 1}_{ab}\right], \\
&V_{ab}(r, r_2, \theta-\theta_2+2\pi)=\frac{(t-l)}{64\pi^2t\epsilon^3}\left[-\frac{t+l}{t}\left(\gamma^1\gamma^t\right)_{ab}-2{\bf 1}_{ab}\right], \\
&V_{ab}(r_2, r, \theta_2-\theta-2\pi)=-\frac{(t-l)}{64\pi^2t\epsilon^3}\left[-\frac{t+l}{t}\left(\gamma^1\gamma^t\right)_{ab}-2{\bf 1}_{ab}\right], \\
&V_{ab}(r, r_2, \theta-\theta_2-2(n-1)\pi)=(-1)^{n-1}\frac{(t-l)}{64\pi^2t\epsilon^3}\left[-\frac{t+l}{t}\left(\gamma^1\gamma^t\right)_{ab}-2{\bf 1}_{ab}\right], \\
&V_{ab}(r_2, r, \theta_2-\theta+2(n-1)\pi)= (-1)^{n} \frac{(t-l)}{64\pi^2t\epsilon^3}\left[-\frac{t+l}{t}\left(\gamma^1\gamma^t\right)_{ab}-2{\bf 1}_{ab}\right].\\
\end{split}
\end{equation}

Let's study the time evolution of $\Delta S^{(n)}_A$ in the next section.
\section{$\Delta S^{(n)}_A$ for Various Local Operators}
In this section, we study the time evolution of $\Delta S^{(n)}_A$ for various operators by the replica trick. Especially, we focus on their behavior in the late time region ($t\gg l$).

\subsection{$\Delta S^{(n)}_A$ for $\psi'_a$}
The locally exited state is given by 
\begin{equation}
\left|\Psi\right\rangle =\mathcal{N} \psi'_a(-t, -l, {\bf x})\left|0\right\rangle.
\end{equation}

The time evolution of $\Delta S^{(n)}_A$ is as follows.
\subsubsection{The Excess of R$\acute{e}$nyi Entanglement Entropies}

$\Delta S^{(n)}_A$ for it is given by 
\begin{equation}
\Delta S^{(n)}_A=\frac{1}{1-n}\log{\left[\frac{\left\langle\psi'^{\dagger}_a(\theta+2(n-1)\pi) \psi' _a(\theta_2+2(n-1)\pi)\cdots\psi'^{\dagger}_a(\theta) \psi' _a(\theta_2)\right\rangle_{\Sigma_n}}{\left\langle\psi'^{\dagger}_a(\theta) \psi' _a(\theta_2)\right\rangle_{\Sigma_1}^n}\right]}.
\end{equation}

When we take the limit $\epsilon \rightarrow 0$, $\Delta S^{(n)}_A$ vanishes in the early time region ($t\le l$). When $t$ is greater than $l$, two diagram in Figure.5  dominantly contribute to $\Delta S^{(n)}_A$. In this region the denominator is given by 
\begin{equation}
\left\langle \psi_a(r, \theta)^{\dagger}\psi_a(r_2, \theta_2)\right\rangle_{\Sigma_1}^n \sim \left(\frac{1}{16\pi^2\epsilon^3}\right)^n.
\end{equation}
Therefore $\Delta S^{(n)}_A$ is given by 
\begin{equation}
\begin{split}
\Delta S^{(n)}_A &=\frac{1}{1-n}\log{\left[A_1+A_2\right]}, \\
A_1&=\left(\frac{t+l}{4t}\right)^n\left(\left(\frac{t-l}{t}\right)(\gamma^t\gamma^1)_{aa}+2\right)^n, \\
A_2&=\left(\frac{t-l}{4t}\right)^n\left(-\left(\frac{t+l}{t}\right)(\gamma^t\gamma^1)_{aa}+2\right)^n.,\\
\end{split}
\end{equation}
where $(\gamma^t \gamma^1)_{aa}$ is real because $\gamma^t \gamma^1$ is a  hermitian matrix\footnote{$-1 \le (\gamma^t \gamma^1)_{aa} \le 1$.}. 
If we take the late time limit ($t \gg l$), $\Delta S^{(n)}_A$ is given by
\begin{equation}
\begin{split}
\Delta S^{(n)}_A &=\frac{1}{1-n}\log{\left[A_1+A_2\right]}, \\
A_1&=\left(\frac{(\gamma^t\gamma^1)_{aa}+2}{4}\right)^n, \\
A_2&=\left(\frac{-(\gamma^t\gamma^1)_{aa}+2}{4}\right)^n.,\\
\end{split}
\end{equation}
where as in Figure.5, $A_1$ and $A_2$ are respectively given by 
\begin{equation}
\begin{split}
&A_1=\frac{\left(V_{aa}(r_2, r, \theta_2 -\theta)\right)^n}{\left\langle \psi'^{\dagger}_a(r, \theta)\psi'_a(r_2, \theta_2)\right\rangle_{\Sigma_1}^n }, \\
&A_2=\frac{(-1)^{n+1}\left(V_{aa}(r_2, r, \theta_2-\theta-2\pi)\right)^{n-1} V_{aa}(r_2, r, \theta_2-\theta+2(n-1)\pi)}{\left\langle \psi'^{\dagger}_a(r, \theta)\psi'_a(r_2, \theta_2)\right\rangle_{\Sigma_1}^n }. \\
\end{split} 
\end{equation}

If we take the Von Neumann limit $n \rightarrow 1$, $\Delta S_A$ is given by
\begin{equation}\label{eega}
\Delta S_A=\frac{1}{4} ((\gamma^t\gamma^1)_{aa}-2) \log (2-(\gamma^t\gamma^1)_{aa})-\frac{1}{4} ((\gamma^t\gamma^1)_{aa}+2) \log ((\gamma^t\gamma^1)_{aa}+2)+\log (4).
\end{equation}
If $(\gamma^t\gamma^1)_{aa}$ vanishes, $\Delta S_A$ is given by $\log{2}$ which is entanglement entropy for the EPR state. The lower value of $\Delta S_A$ in (\ref{eega}) is given by $\log{4/3^{\frac{3}{4}}}$ ($(\gamma^t\gamma^1)_{aa} =1$ or $-1$).

\subsubsection{Reduced Density Matrix}
If we identify $A_1$ and $A_2$ as the diagonal components of $\rho_A^n$ respectively, 
it is expected that the density matrix for this state is given by 
\begin{equation}\label{phi}
\rho_A =\frac{1}{4}
\begin{pmatrix}
(\gamma^t\gamma^1)_{aa}+2& 0 \\
0 & -(\gamma^t\gamma^1)_{aa}+2 \\
\end{pmatrix},
\end{equation}
where $\tr_A \rho_A$ and each diagonal components of $\rho_A$ are positive.

It is expected that the density matrix for $\left|\Psi\right\rangle =\mathcal{N}\bar{\psi}'_a\left|0\right\rangle$ is given by 

\begin{equation}\label{rdphi}
\rho_A =\frac{1}{4}
\begin{pmatrix}
(\gamma^1\gamma^t)_{aa}+2& 0 \\
0 & -(\gamma^1\gamma^t)_{aa}+2 \\
\end{pmatrix}
.
\end{equation}
\begin{figure}[h]
 \begin{minipage}{0.5\hsize}
  \begin{center}
   \includegraphics[width=65mm]{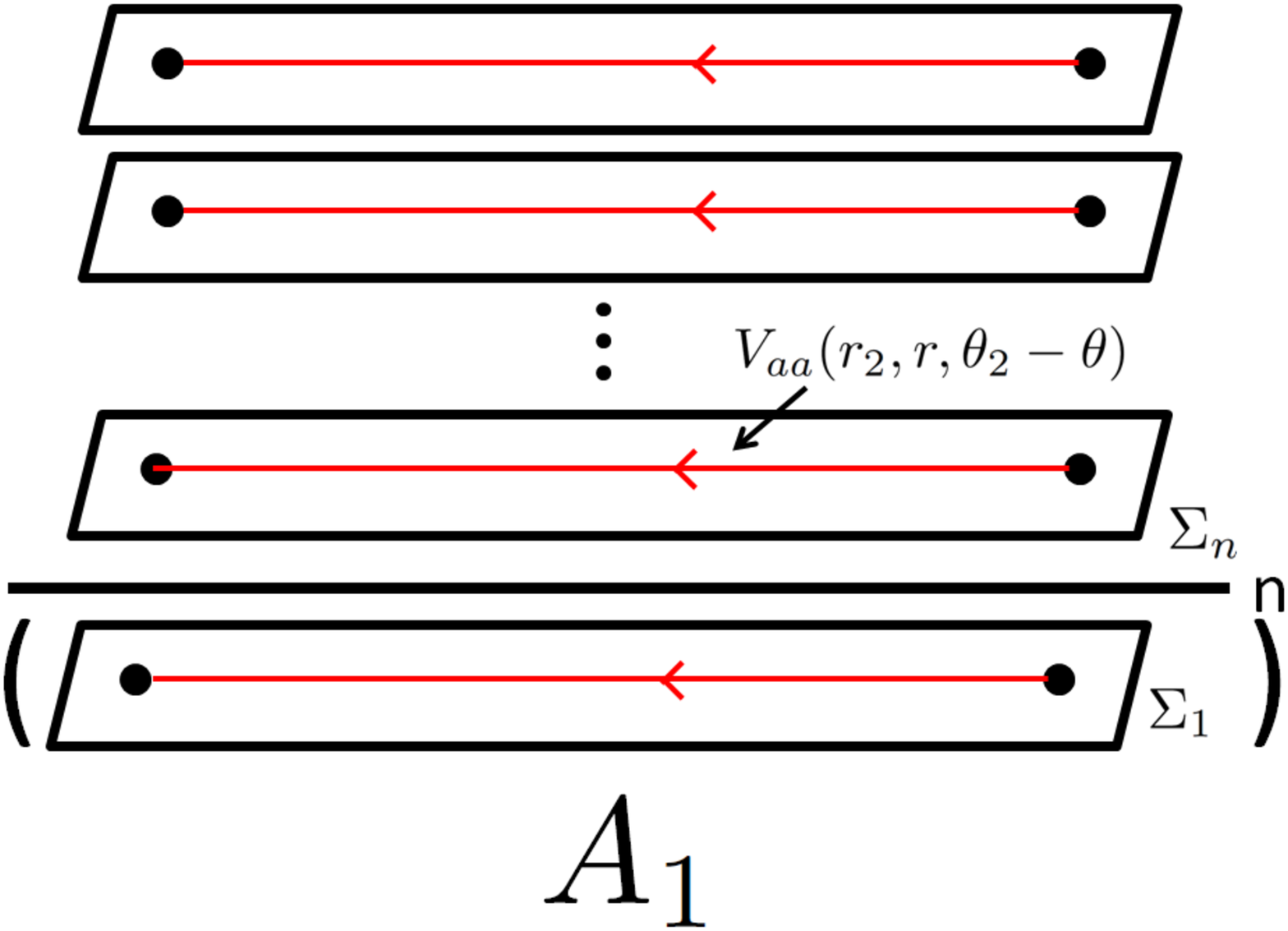}
  \end{center}
 \end{minipage}
  \begin{minipage}{0.30\hsize}
  \begin{center}
   \includegraphics[width=65mm]{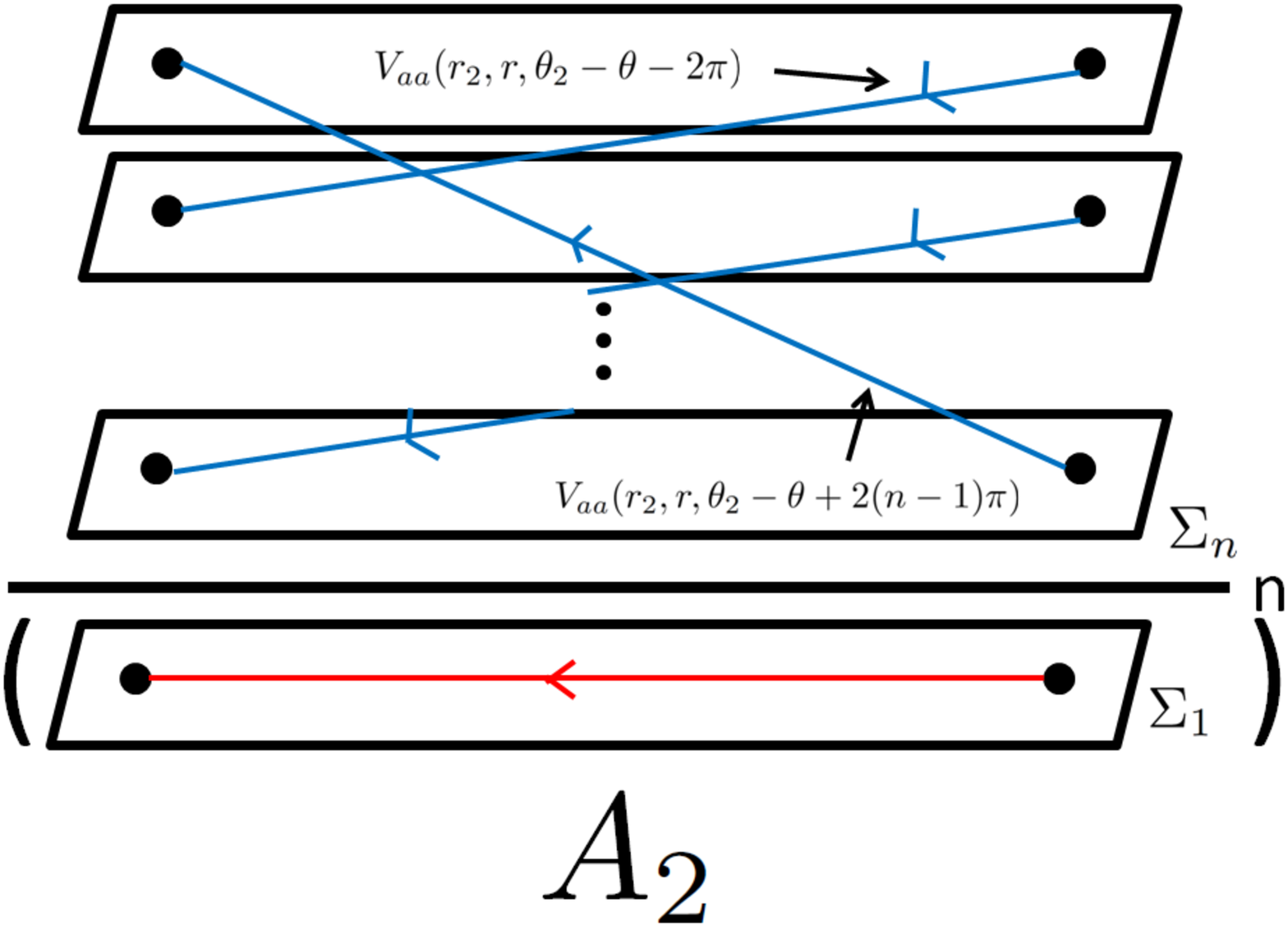}
  \end{center}
 \end{minipage}
 \caption{The schematic description of $A_1$ and $A_2$.}\label{A1A2}
\end{figure}
\subsection{$\Delta S^{(n)}_A$ for $\bar{\psi}'\psi'$}
A locally excited state is given by 
\begin{equation}\label{ins}
\left|\Psi\right\rangle = \mathcal{N}\bar{\psi}'\psi'(-t, -l, {\bf x})\left|0\right\rangle, 
\end{equation}
which is invariant for $SL(2,{\bf C})$ transformation.

Let's study the time evolution of $\Delta S^{(n)}_A$.
\subsubsection{The Excess of (R$\acute{e}$nyi) Entanglement Entropy}
$\Delta S^{(2)}_A$ for (\ref{ins}) is given by 
\begin{equation}
\Delta S^{(2)}_A =- \log{\left[\frac{\left\langle\bar{\psi}\psi (\theta+2\pi) \bar{\psi}\psi ( \theta_2+2\pi) \bar{\psi}\psi (\theta) \bar{\psi}\psi (\theta_2) \right\rangle_{\Sigma_2}}{\left\langle\bar{\psi}\psi(\theta) \bar{\psi}\psi ( \theta_2)\right\rangle_{\Sigma_1}^2}\right]}
\end{equation}

In the early time region ($t\le l$), $\Delta S^{(2)}_A$ vanishes when we take the limit $\epsilon \rightarrow 0$. On the other hand, in the  region $t>l$, $\Delta S^{(2)}_A$ is given by
\begin{equation}
\begin{split}
&\Delta S^{(2)}_A=-\log{[A_1+A_2+A_3]}, \\
&A_1=\left[\frac{c(t+l)^4}{2^4c~ t^2(t^2-l^2)}\left[ \left(\frac{2t-l}{t}\right)^2-1\right]\right]^2, \\
&A_2=\left[\frac{c(t-l)^4}{2^4c~ t^2(t^2-l^2)}\left[ \left(\frac{2t+l}{t}\right)^2-1\right]\right]^2, \\
&A_3=2c\left[\left(\frac{t^2-l^2}{2^4t^2c}\right)^2\left[\left(\frac{5t^2-l^2}{t^2}\right)^2+16\right]\right], \\
\end{split}
\end{equation}
where $c =4$.

If we take the limit $t\rightarrow \infty$, $\Delta S^{(2)}_A$ is given by 

\begin{equation}
\Delta S^{(2)}_A =\log{\left[\frac{2^9}{77}\right]}
\end{equation}

If $t\le l$, for arbitrary $n$, $\Delta S^{(n)}_A$ vanishes.
In the region $t>l$, $\Delta S^{(n)}_A$ is given by
\begin{equation}
\Delta S^{(n)}_A =\frac{1}{1-n}\log{\left[A_1+A_2+A_3\right]},
\end{equation}
where $A_1, A_2, A_3$ are given by
\ba
A_1&=&\left[
{c(t+l)^4\over 2^4c t^2(t^2-l^2)}
\left[
\left({2t-l\over t}\right)^2-1
\right]
\right]^n \cr
A_2&=&\left[
{c(t-l)^4\over 2^4c t^2(t^2+l^2)}
\left[
\left({2t-l\over t}\right)^2-1
\right]
\right]^n \cr
A_3&=&
2c
\left(
{t^2-l^2\over 2^4 t^2 c}
\right)^n
\left[
\sum_{k\in 2\mathbb{Z}}^{k\le n}
{}_nC_k
4^k \left({5t^2-l^2\over t^2}\right)^{n-k}
\right].
\ea

If we take the late time limit $t \rightarrow \infty$, $\Delta S^{(n)}_A$ is given by
\begin{equation}
\Delta S^{(n)}_A = \frac{1}{1-n} \log{\left[2\left(\frac{12}{64}\right)^n+8 \left(\frac{1}{64}\right)^n \sum_{k\in 2\mathbb{Z}}^{k\le n}{ }_nC_k4^k 5^{n-k}
\right]}.
\end{equation}

\subsubsection{Reduced Density Matrix}
By using results up to here, we are able to guess the reduce density matrix for $\left|\Psi \right\rangle =\mathcal{N}\bar{\psi}'\psi'$ as follows. In the limit $\epsilon \rightarrow 0$, only four diagrams in Figure.6 can contribute to $\Delta S^{(n)}_A$. We define $A_1,A_2, A$ by 
\begin{equation}
\begin{split}
&A_1 = \frac{12}{64}, \\
&A_2 = \frac{12}{64}, \\
&A =\frac{1}{64}
\begin{pmatrix}
5 & 4 \\
4 & 5
\end{pmatrix}
.
\end{split}
\end{equation}
In the replica trick, $(A_1)^n$, $(A_2)^n$ and $8\cdot tr(A^n)$ respectively correspond to the red-lined diagram, blue lined diagram and green lined diagrams in Figure.6. Therefore it is expected that the reduced density matrix for this state is given by the $10 \times 10$ matrix,
\begin{equation}
\rho_A =\frac{1}{64}\begin{pmatrix}
12 & 0 & 0 & 0 & 0 &0 \\
0 & \tilde{A} & 0 & 0 & 0 &0  \\
0 & 0 & \tilde{A} & 0 & 0 & 0  \\
0 & 0 & 0 & \tilde{A} & 0 & 0   \\
0 & 0 & 0 & 0 & \tilde{A} & 0   \\
0 & 0 & 0 & 0 & 0 & 12 \\
\end{pmatrix}
,
\end{equation}
where $\tilde{A}$ is given by the $2\times 2$ matrix,
\begin{equation}
\tilde{A}=
\begin{pmatrix}
5 & 4 \\
4 & 5
\end{pmatrix}
.
\end{equation}
If we diagonalize the matrix $\tilde{A}$, it is given by
\begin{equation}
\tilde{A}=
\begin{pmatrix}
9 & 0 \\
0 &1
\end{pmatrix}
.
\end{equation}

For any $n$, (R$\acute{e}$nyi) entanglement entropy is given by 
\begin{equation}
\Delta S^{(n)}_A =\frac{1}{1-n}\log{\left[\frac{2\cdot12^n+4\cdot9^n+4}{2^{6n}}\right]}
\end{equation}
If we take the von Neumann entropy limit $n\rightarrow 1$, entanglement entropy is given by 
\begin{equation}
\Delta S_A =\frac{3}{4}\log{\left(\frac{128}{9}\right)}.
\end{equation}
Min entropy is given by 
\begin{equation}\label{min1}
\Delta S^{(\infty)}_A =\log{\left(\frac{16}{3}\right)}.
\end{equation}
(R$\acute{e}$nyi) entanglement entropy for this state monotonically decreases when the replica number $n$ increases. Therefore we can consider $\Delta S^{(\infty)}_A$ as the lower bound of (R$\acute{e}$nyi) entanglement entropies. 

If we take the large $N$ limit in the $U(N)$ or $SU(N)$ free massless fermionic field theory, the number of diagrams which can dominantly contribute to $\Delta S^{(n)}_A$ decreases.
Because the number of trace in the green-lined diagram is less than that in the others, the blue-lined and red-lined diagram dominantly contribute to $\Delta S^{(n)}_A$ in the large $N$ limit.
Therefore it is given by 
\begin{equation}
\Delta S^{(n\ge2)}_A=\frac{2n-1}{n-1}\log{2}+\frac{n}{1-n}\log{\left(\frac{3}{4}\right)}.
\end{equation}
After that, if we take $n\rightarrow \infty$ limit, Min entropy is given by 
\begin{equation}
\Delta S^{(\infty)}_A =\log{\left(\frac{16}{3}\right)}.
\end{equation}
This result describes that Min entropy is same as (\ref{min1}) even if we take the large $N$ limit.

\begin{figure}[h]
\centering
 \begin{minipage}{0.2\hsize}
  \begin{center}
   \includegraphics[width=30mm]{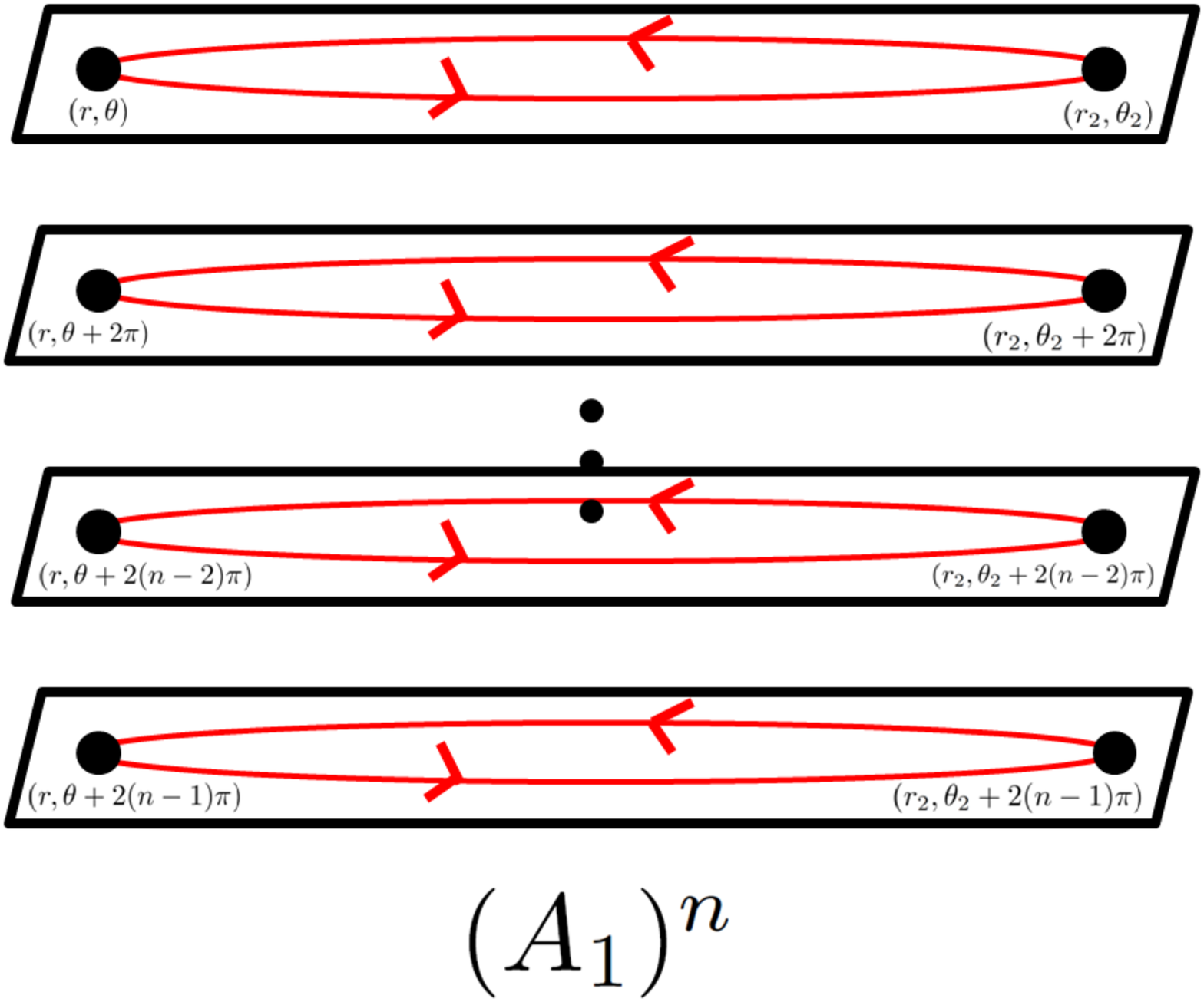}
  \end{center}
 \end{minipage}
  \begin{minipage}{0.2\hsize}
  \begin{center}
   \includegraphics[width=30mm]{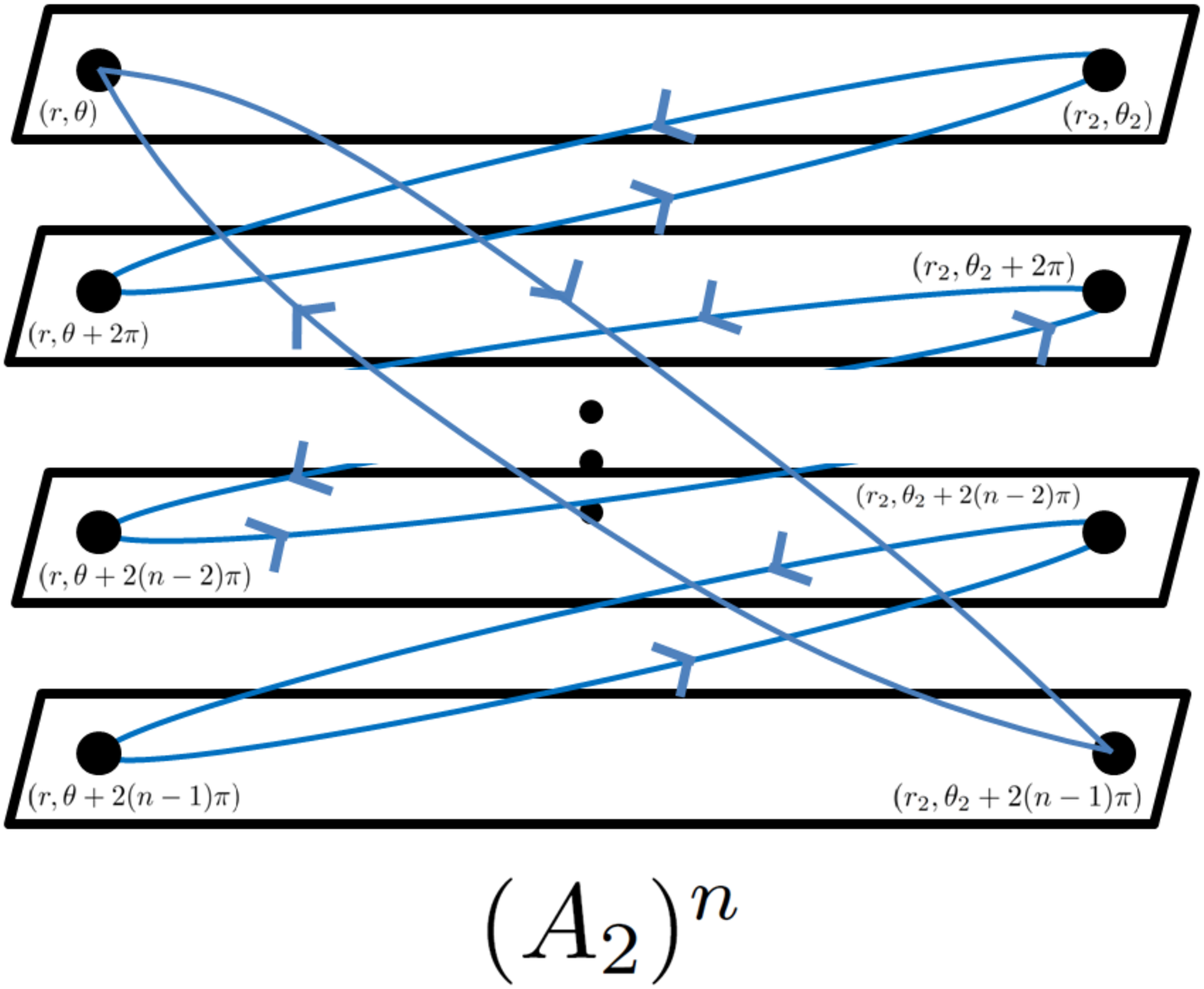}
  \end{center}
 \end{minipage}
 \begin{minipage}{0.2\hsize}
  \begin{center}
   \includegraphics[width=30mm]{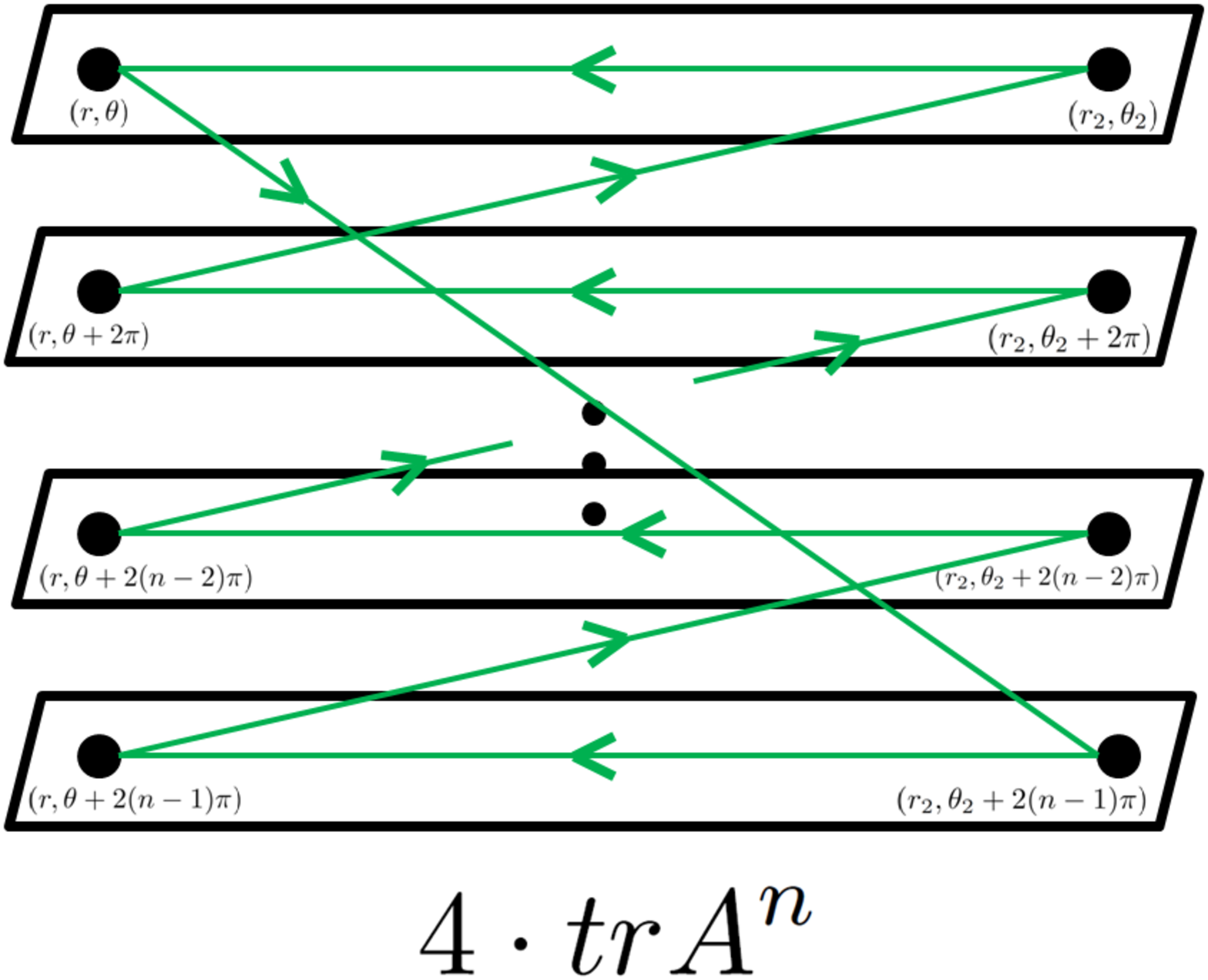}
  \end{center}
 \end{minipage}
  \begin{minipage}{0.2\hsize}
  \begin{center}
   \includegraphics[width=30mm]{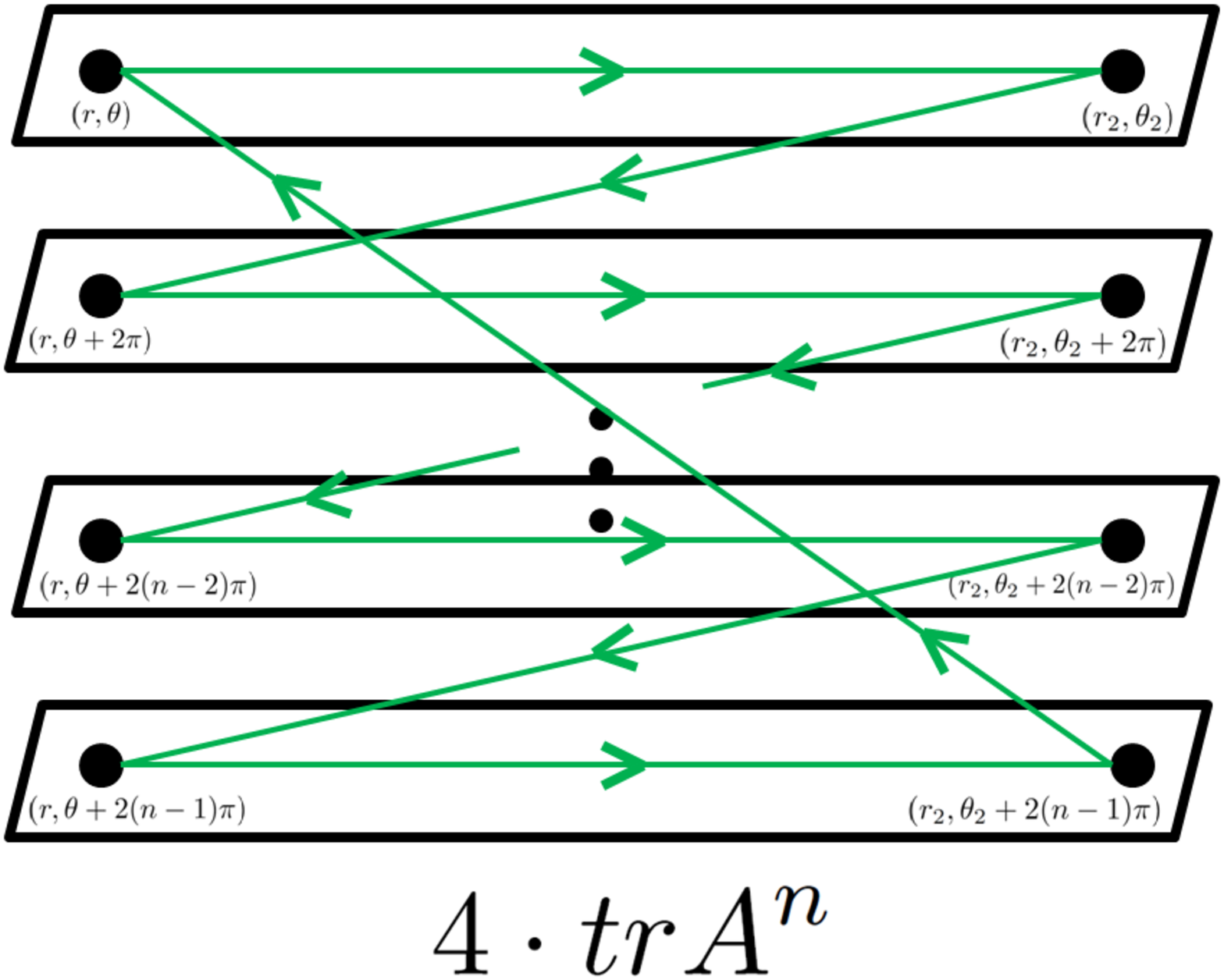}
  \end{center}
 \end{minipage}
 \caption{The schematic explanation of the correspondence between diagrams and components of a reduce density matrix. }
\end{figure}

\subsection{$\Delta S^{(n)}_A$ for $\psi'^{\dagger}\psi'$}
A locally excited state is given by 
\begin{equation}\label{uns}
\left|\Psi\right\rangle = \mathcal{N}\psi'^{\dagger} \psi'(-t,-l, {\bf x})\left|0\right\rangle, 
\end{equation}
which is variant under the $SL(2,{\bf C})$ transformation.

\subsubsection{The Excess of 
 (R$\acute{e}$nyi) Entanglement Entropy}
$\Delta S^{(2)}_A$ for the state in (\ref{uns}) is given by
\begin{equation}
\begin{split}
\Delta S^{(2)}_A =-\log{\left[\frac{\left\langle \psi'^{\dagger}\psi'(\theta+2\pi) \psi'^{\dagger}\psi'(\theta_2+2\pi) \psi'^{\dagger}\psi'(\theta) \psi'^{\dagger}\psi'(\theta_2)\right\rangle_{\Sigma_2}}{\left\langle \psi'^{\dagger}\psi'(\theta) \psi'^{\dagger}\psi'(\theta_2)\right\rangle_{\Sigma_1}^2 }\right]}.
\end{split}
\end{equation}

If we take the limit $\epsilon \rightarrow 0$, $\Delta S^{(2)}_A$ vanishes in the early time region ($t \le l$). In the region $t>l$, $\Delta S^{(2)}_A$ nontrivially grows as follows,
\begin{equation}
\begin{split}
\Delta S^{(2)}_A &=- \log{[A_1+A_2+A_3+A_4]}, \\
A_1&=\frac{(t+l)^4}{4^6t^4}\left(16+4\left(\frac{t-l}{t}\right)^2\right)^2, \\
A_2&=\frac{(t-l)^4}{4^6t^4}\left(16+4\left(\frac{t+l}{t}\right)^2\right)^2, \\
A_3&=\frac{(t^2-l^2)^2}{4^6t^4}\left(4\left(\frac{t^2-l^2}{t^2}-4\right)^2+4\times\frac{4l^2}{t^2}\right), \\
A_4&=\frac{(t^2-l^2)^2}{4^6t^4}\left(4\left(\frac{t^2-l^2}{t^2}-4\right)^2+4\times\frac{4l^2}{t^2}\right). \\
\end{split}
\end{equation}
If we take the late time limit ($t \gg l$), $\Delta S^{(2)}_A $ is given by 
\begin{equation}
\begin{split}
\Delta S^{(2)}_A=-\log{\left[\frac{20^2}{4^6}+\frac{20^2}{4^6}+\frac{4\cdot 3^2}{4^6}+\frac{4\cdot 3^2}{4^6}\right]}.
\end{split}
\end{equation}

If $t\le l$, for arbitrary $n$, $\Delta S^{(n)}_A$ vanishes.
In the region $t>l$, $\Delta S^{(n)}_A$ is given by
\begin{equation}
\Delta S^{(n)}_A =\frac{1}{1-n}\log{\left[A_1+A_2+A_3\right]},
\end{equation}
where $A_1, A_2, A_3$ are given by
\ba
A_1&=&\left[
{c(t+l)^2\over 2^4c t^2}
\left[
\left({t-l\over t}\right)^2+4
\right]
\right]^n \cr
A_2&=&\left[
{c(t-l)^2\over 2^4c t^2}
\left[
\left({t+l\over t}\right)^2+4
\right]
\right]^n \cr
A_3&=&
2c
\left(
{t^2-l^2\over 2^4 t^2 c}
\right)^n
\left[
\sum_{k\in 2\mathbb{Z}}^{k\le n}
{}_nC_k
\left(\frac{4l}{t}\right)^k \left({3t^2+l^2\over t^2}\right)^{n-k}
\right].
\ea

If we take the late time limit $t \rightarrow \infty$, $\Delta S^{(n)}_A$ is given by
\begin{equation}\label{runs}
\Delta S^{(n)}_A =\frac{1}{1-n}\log{\left[2\left(\frac{20}{64}\right)^n+8\left(\frac{3}{64}\right)^n\right]}.
\end{equation}

\subsection{Reduced Density Matrix}
We define $B_1, B_2, B $ by 
\begin{equation}
\begin{split}
&B_1=\frac{20}{64}, \\
&B_2=\frac{20}{64}, \\
&B=\frac{1}{64}
\begin{pmatrix}
3 & 0 \\
0 & 3 \\
\end{pmatrix}
.
\end{split}
\end{equation}

In the replica trick, $(B_1)^n$, $(B_2)^n$ and $8\cdot tr(B^n)$ respectively correspond to the red dashed lined diagram, blue dashed lined diagram and green dashed lined diagrams in Figure.7. Therefore it is expected that the reduced density matrix for this state is given by the $10 \times 10$ matrix,
\begin{equation}
\rho_A =\frac{1}{2^6}\begin{pmatrix}
20 & 0 & 0 & 0 & 0 &0 \\
0 & \tilde{B} & 0 & 0 & 0 &0  \\
0 & 0 & \tilde{B} & 0 & 0 & 0  \\
0 & 0 & 0 & \tilde{B} & 0 & 0   \\
0 & 0 & 0 & 0 & \tilde{B} & 0   \\
0 & 0 & 0 & 0 & 0 & 20 \\
\end{pmatrix}
,
\end{equation}
where $\tilde{B}$ is given by $2\times 2$ matrix,
\begin{equation}
\tilde{B} =
\begin{pmatrix}
3 & 0 \\
0 & 3
\end{pmatrix}
.
\end{equation}

If we take Von Neumann entropy limit $n\rightarrow 1$ in $(\ref{runs})$ entanglement entropy is given by 
\begin{equation}
\Delta S_A =\log{16}+\frac{3}{4}\log{2}-\frac{3}{8}\log{3}-\frac{5}{8}\log{5}.
\end{equation}
Although the local operator $\psi'^{\dagger}\psi'$ is not invariant under $SL(2, {\bf C})$ transformation, $\Delta S^{(n)}_A$ is invariant under this transformation.

If we take the limit $n\rightarrow \infty$, it is given by
\begin{equation}\label{spp}
\Delta S^{(\infty)}_A=\log{\left(\frac{16}{5}\right)}
\end{equation}

If we take the large $N$ limit in the $U(N)$ or $SU(N)$ free massless fermionic field theory, the number of diagrams which can contribute to $\Delta S^{(n)}_A$ decreases.
Because the number of trace in green dashed lined diagram is less than that in the others, blue dashed lined and red dashed lined diagram dominantly contribute to $\Delta S^{(n)}_A$ in the large $N$ limit.
Therefore it is given by 
\begin{equation}
\Delta S^{(n)}_A =\frac{1}{1-n}\log{2}+\frac{n}{1-n}\log{\left(\frac{5}{16}\right)}.
\end{equation}
If we take $n \rightarrow \infty$ limit, $\Delta S^{(\infty)}_A$ is given by 
\begin{equation}
\Delta S^{(\infty)}_A = \log{\left(\frac{16}{5}\right)}.
\end{equation}
Even if we take the large $N$ limit, the lower bound of $\Delta S^{(n)}_A$ agrees with that in (\ref{spp}) similarly to the result in the previous subsection.

\begin{figure}[h]
\centering
 \begin{minipage}{0.2\hsize}
  \begin{center}
   \includegraphics[width=30mm]{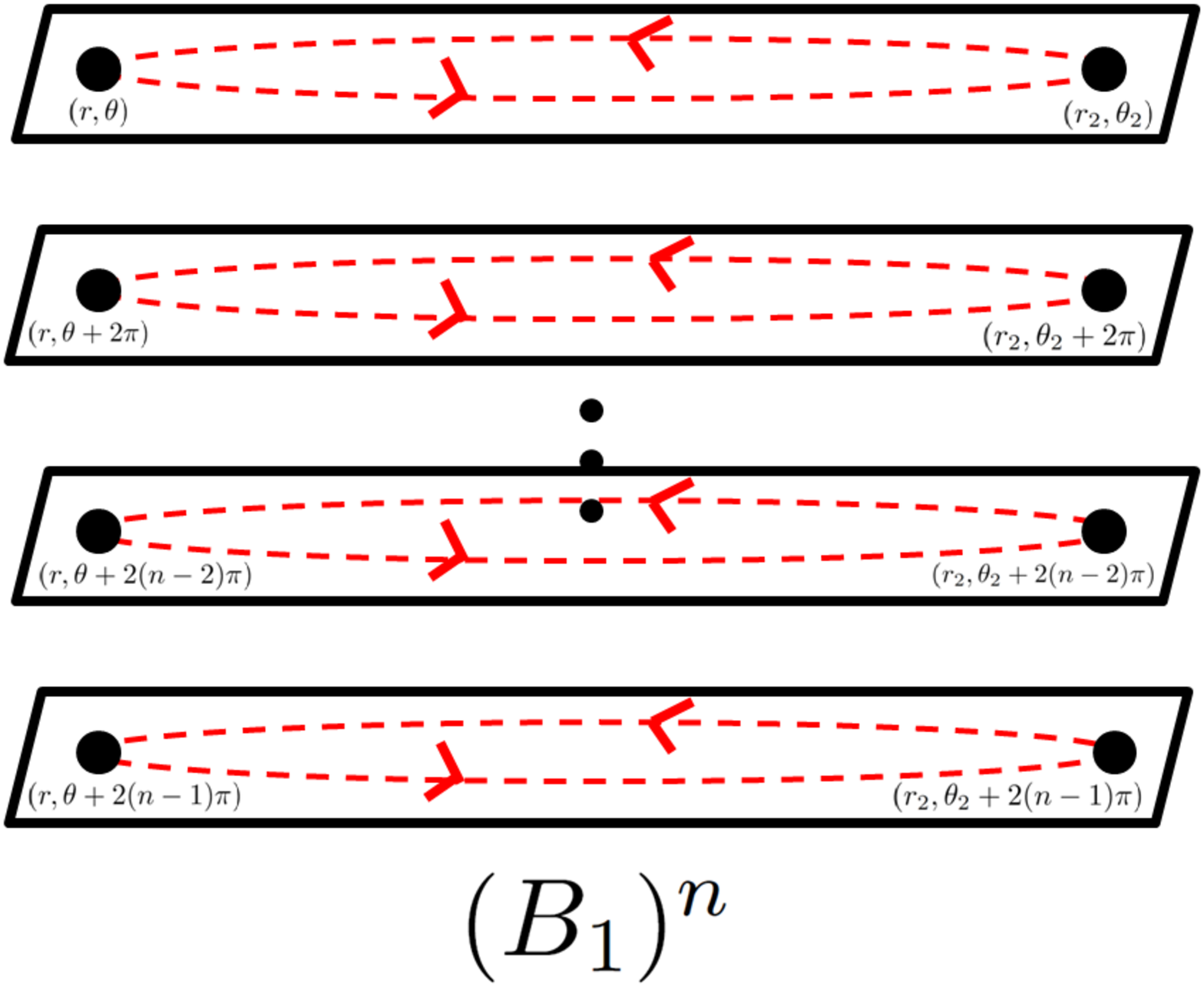}
  \end{center}
 \end{minipage}
  \begin{minipage}{0.2\hsize}
  \begin{center}
   \includegraphics[width=30mm]{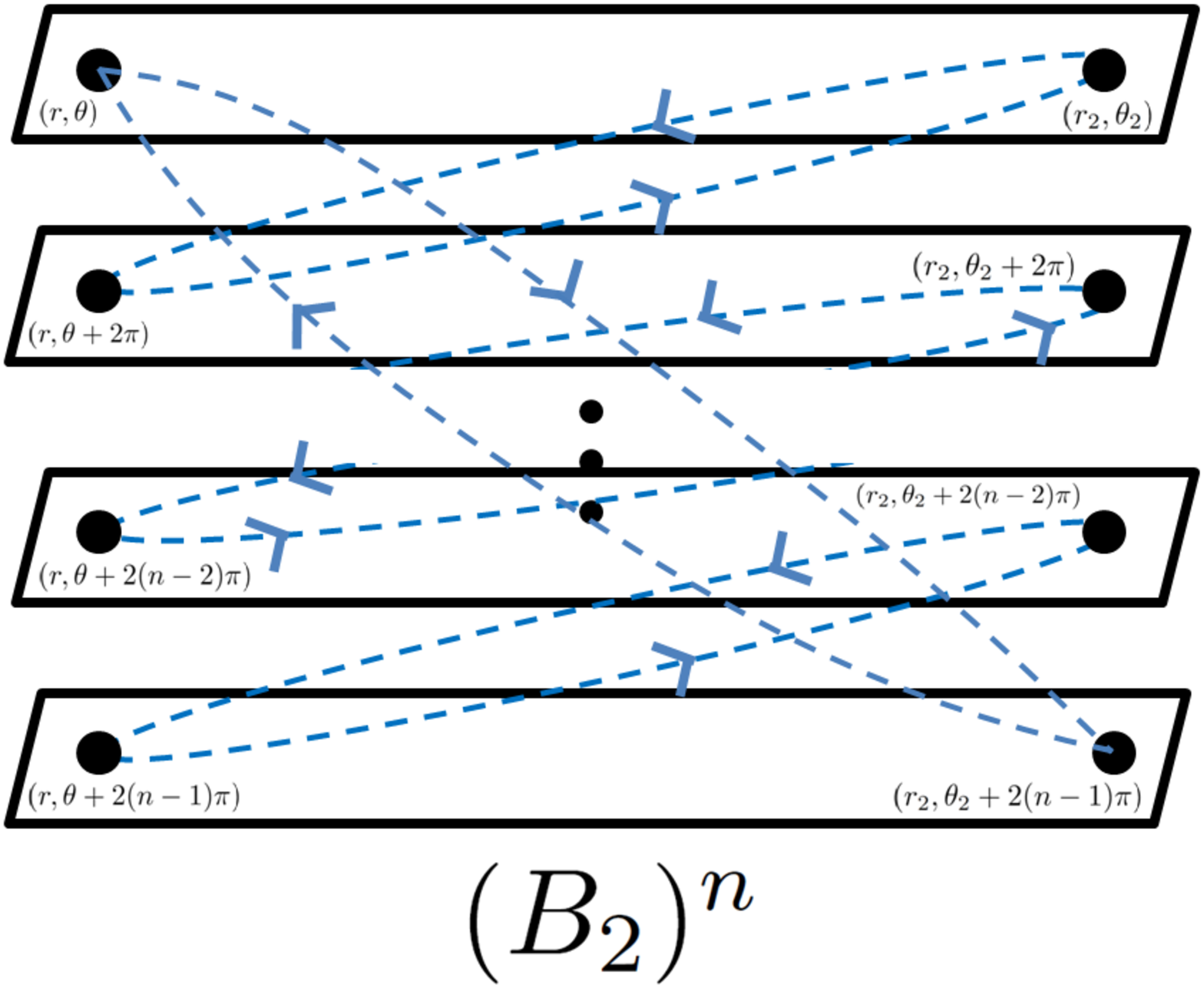}
  \end{center}
 \end{minipage}
 \begin{minipage}{0.2\hsize}
  \begin{center}
   \includegraphics[width=30mm]{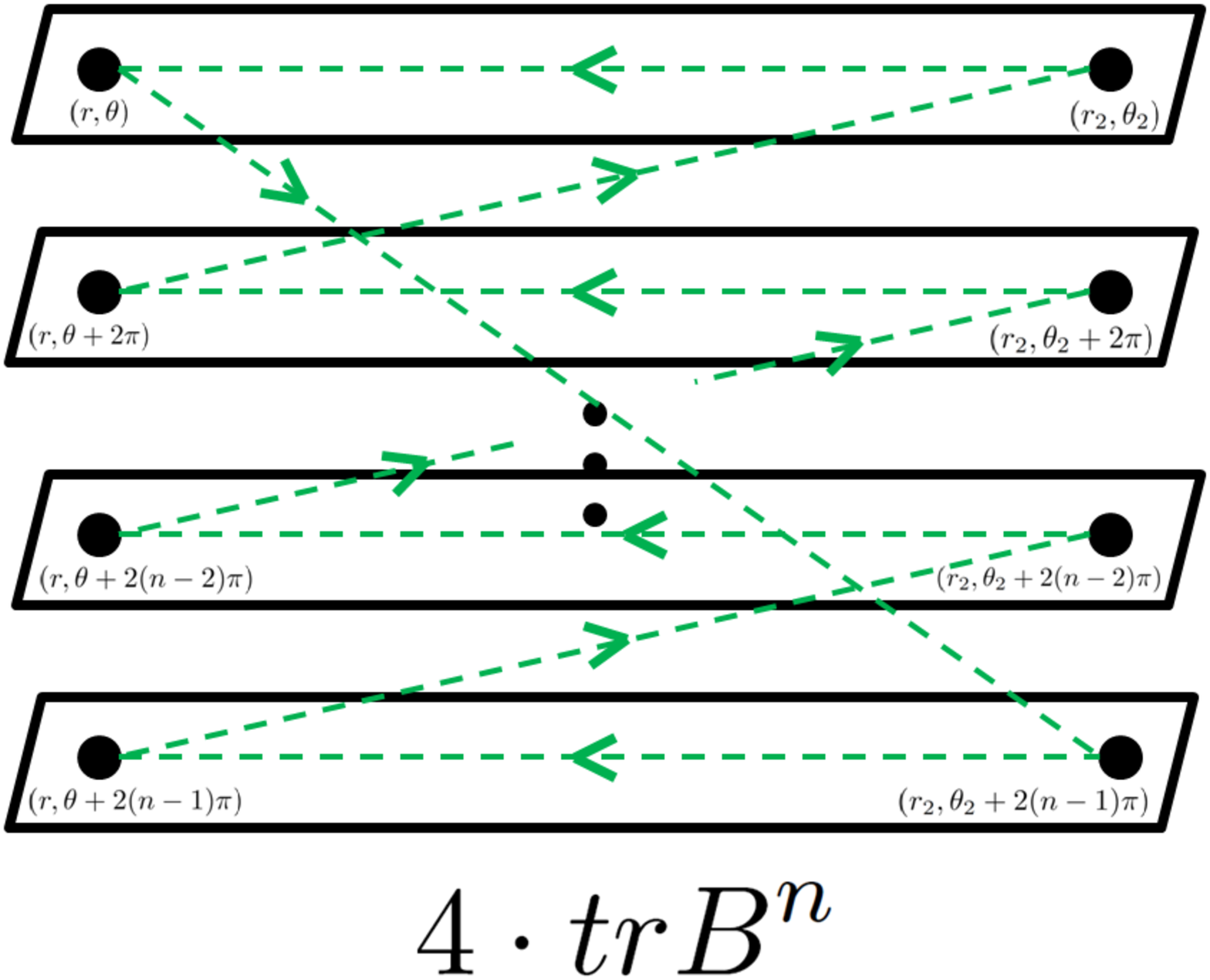}
  \end{center}
 \end{minipage}
  \begin{minipage}{0.2\hsize}
  \begin{center}
   \includegraphics[width=30mm]{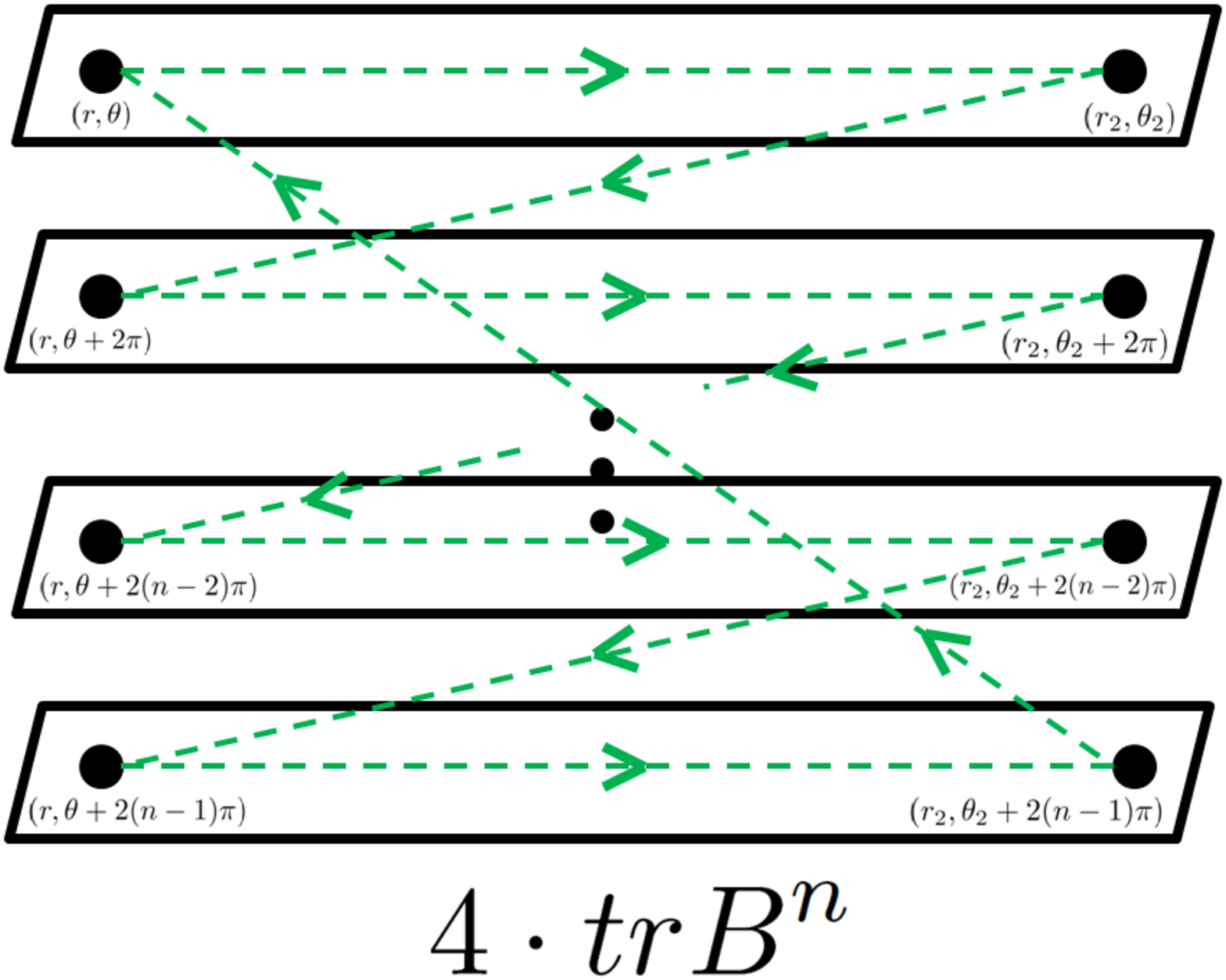}
  \end{center}
 \end{minipage}
 \caption{The schematic explanation of the correspondence between diagrams and components of a reduce density matrix. }
\end{figure}

\subsection{Spin dependence}
Here we discuss the spin dependence of the reduced density matrices in (\ref{phi}) and (\ref{rdphi}). Their components have $\left(\gamma^t\gamma^1\right)_{aa}$. Therefore they depend on the direction of the spin. These spin dependences can be understood as follows. To explain why matrix components depend on the spin, we diagonalize $\gamma^t\gamma^1$ and it is given by
\begin{equation}\label{gtg1}
\begin{split}
\gamma^t\gamma^1 =\begin{pmatrix}
1&0&0&0\\
0&-1&0&0 \\
0&0&1&0 \\
0&0&0&-1 \\
\end{pmatrix}.
\end{split}
\end{equation}  
We can derive $\left(p_0+\gamma^t\gamma^ip_i\right)\Phi(p)=0$ from the equation of motion for $\Phi(p)$. The modes propagating along $x^1$ direction ($p_1\neq0, p_2=0, p_3=0$) obey 
\begin{equation}\label{eom}
\left(p_0+p_1\gamma^t\gamma^1\right)\Phi(p)=
\begin{pmatrix}
p_0+p_1&0&0&0\\
0&p_0-p_1&0&0 \\
0&0&p_0+p_1&0 \\
0&0&0&p_0-p_1 \\
\end{pmatrix}
\begin{pmatrix}
\Phi^{+}_1(p) \\
\Phi^{-}_2(p) \\
\Phi^{+}_3(p) \\
\Phi^{-}_4(p) \\
\end{pmatrix}.
=0.
 \end{equation}
 
The equation in (\ref{eom}) describes that that (anti-)particles can propagate along only the left or right direction as follow.  
An anti-particle $\Phi^{+}(p)$ $(\gamma^t\gamma^1\Phi^{+}(p)= \Phi^{+}(p))$ can not propagate in the right direction parallel to $x^1$ axis ($x^1>0$). On the other hand, the anti-particle $\Phi^{-}(p)$ $(\gamma^t\gamma^1\Phi^{-}(p)=- \Phi^{-}(p))$  is not able to propagate in the left direction ($x^1<0$).
Therefore if we define a locally excited state by acting a component of $\psi$ or $\bar{\psi}$ on the ground state, their reduced density matrices depend on the representation of gamma matrices as in (\ref{phi}) and (\ref{rdphi}). 
For example, we choose the basis which diagonalizes $\gamma^t\gamma^1$ as in (\ref{gtg1}) and $\psi_{1}$ acts on the ground state. This local operator creates an anti-particle and it propagates with time. At the late time ($t\gg l$), it is necessarily included in the region A or B. As explained above, it is not able to propagate in the right direction parallel to $x^1$ axis. Therefore it is included in $A$ ($B$) with the probability which is bigger than $\frac{1}{2}$ (less than $\frac{1}{2}$).  The reduced density matrix in (\ref{phi}) is given by
\begin{equation}\label{oof}
\begin{split}
\rho_A =\frac{1}{4}
\begin{pmatrix}
3 &0 \\
0& 1 \\
\end{pmatrix}
.
\end{split}
\end{equation}    
The result in (\ref{oof}) describes that the antiparticle which created by $\psi_1$ are included in A (B) with probability $\frac{3}{4}$ ($\frac{1}{4}$)\footnote{If we act $\psi_1+\psi_2$ on the ground state, the late time value of $\Delta S^{(2)}_A$ is given by $\log{2}$ as we expected}\footnote{ If you choose the bases which diagonalize $\gamma^t \gamma^1$ in 2d massless free fermionic field theory, a component of $\psi$ ($\bar{\psi}$) is able to propagate in the only left or right direction parallel to $x^1$. Therefore $\Delta S^{(n)}_A$ vanish for the locally excited states generated by acting a component of $\psi$ ($\bar{\psi}$) on the ground state.}. If we choose a basis in which $\left(\gamma^t \gamma^1\right)_{11}$ is given by $0$, $\Delta S^{(n)}_A$ is given by $\log{2}$.


\section{Quasi-Particle Interpretation}
In \cite{m1, m2, m3}, we have interpreted the late time value of $\Delta S^{(n)}_A$ in free massless scalar field theories in terms of entanglement between quasi-particles.

In free fermionic field theory, it is expected that the late time values of $\Delta S^{(n)}_A$ can be interpreted in terms of entanglement between quasi-particles.
Therefore we decompose local operators into left moving modes and right moving modes, 
\begin{equation}\label{DQ}
\begin{split}
\psi_a&=\psi_a^{L\dagger}+\psi_a^{R\dagger}+\phi^{L}_a+\phi^{R}_a, \\
\psi_a^{\dagger}&=\psi_a^{L}+\psi_a^{R}+\phi^{L\dagger}_a+\phi^{R\dagger}_a, \\
\bar{\psi}_a&=\bar{\psi}^{L\dagger}_a+\bar{\psi}^{R\dagger}_a+i\left(\psi^L\gamma^t\right)_a+i\left(\psi^R\gamma^t\right)_a,
\end{split}
\end{equation}
where $\bar{\psi}^{K\dagger}$ is defined by
\begin{equation}
\begin{split}
\bar{\psi}^{L\dagger}_a \equiv i\left(\phi^{L\dagger}\gamma^t\right)_a,\\
\bar{\psi}^{R\dagger}_a \equiv i\left(\phi^{R\dagger}\gamma^t\right)_a.\\
\end{split}
\end{equation}

The left moving mode and right moving mode of the anti-particle and particle are defined as follows. We decompose $\psi, \psi^{\dagger}$ into the momentum modes. 
Their right moving modes (left moving modes) are defined by the sum of the momentum modes whose $\frac{p_0}{p_1}$ is positive ($\frac{p_0}{p_1}$ is negative). As we explained in the previous section, the number of momentum modes which the right moving modes (left moving modes) have depends on the choice of spin's direction. Therefore instead of ordinary anti-commutation relationships ($\{\Phi^K_a, \Phi^{K'\dagger}_b\}=\delta_{K K'}\delta_{ab}$), we impose the following exotic anti-commutation relationship on the particle and anti-particle,

\begin{equation}\label{dac}
\begin{split}
\{\phi^R_a, \phi^{R\dagger}_b\}&=\left(\delta_{ab}-\frac{1}{2}\left(\gamma^t\gamma^1\right)_{ab}\right), \\
\{\phi^L_a, \phi^{L\dagger}_b\}&=\left(\delta_{ab}+\frac{1}{2}\left(\gamma^t\gamma^1\right)_{ab}\right), \\
\{\psi^{R\dagger}_a, \psi^{R}_b\}&=\left(\delta_{ab}-\frac{1}{2}\left(\gamma^t\gamma^1\right)_{ab}\right), \\
\{\psi^{L\dagger}_a, \psi^{L}_b\}&=\left(\delta_{ab}+\frac{1}{2}\left(\gamma^t\gamma^1\right)_{ab}\right), \\
\{\bar{\psi}^R_a, \bar{\psi}^{R\dagger}_b\}&=\left(\delta_{ab}-\frac{1}{2}\left(\gamma^1\gamma^t\right)_{ab}\right), \\
\{\bar{\psi}^L_a, \bar{\psi}^{L\dagger}_b\}&=\left(\delta_{ab}+\frac{1}{2}\left(\gamma^1\gamma^t\right)_{ab}\right). \\
\end{split}
\end{equation}

The vacuum $\left|0\right\rangle=\left|0\right\rangle_L\otimes\left|0\right\rangle_R$ is defined by
\begin{equation}
\phi^K_a\left|0\right\rangle =\psi^K_a\left|0\right\rangle =\bar{\psi}^K_a\left|0\right\rangle=0,
\end{equation}
where $K=L, R$.
The number of momentum modes which the left and right moving modes have depends on the choice of $\gamma^t \gamma^1$ (the choice of the spin's direction). Therefore, it is reasonable that the anti-commutation for them is given by that in (\ref{dac})\footnote{However we find this exotic anti-commutation relationship heuristically.}. We claim that the exotic anti-commutation relationship in (\ref{dac}) can be applied to $\Delta S^{(n)}_A$ in $4$ dimensional free massless fermionic field theory. However in $d(\neq4)$ dimensional spacetime, that in (\ref{dac}) should be deformed. For example, in $2$ dimensional spacetime the reduced density matrix for the state excited by a local operator constructed of chiral  and anti-chiral operators should become 
\begin{equation}
\rho_A = \frac{1}{2}\begin{pmatrix}
1+(\gamma^t \gamma^1)_{aa}& 0 \\
0& 1-(\gamma^t \gamma^1)_{aa} \\
\end{pmatrix}.
\end{equation}
Therefore the anti-commutation relation in (\ref{dac}) should change to another one.

This way, we can construct reduced density matrices which agree with those which are obtained by the replica trick as we will see later. In following subsections we will compute them under this decomposition in various examples in $4$ dimensional free massless fermionic field theory.

\subsection{Various Examples}
Here we compute reduced density matrices for various locally excited states under the quasi-particle interpretation.

\subsubsection{$\rho_A$ for $\psi_{a}$}

A locally excited states is given by 
\begin{equation}
\left |\Psi \right \rangle = \mathcal{N} \psi_a\left|0\right\rangle,
\end{equation}
which is variant under the $SL(2, {\bf C})$ transformation.
Under the decomposition in (\ref{DQ}), this state is given by 
\begin{equation}
\left|\Psi\right\rangle = \frac{1}{\sqrt{2}}\left[\psi^{L\dagger}_a\left|0\right\rangle_L\otimes\left|0\right\rangle_R+\left|0\right\rangle_L\otimes \psi^{R\dagger}_a\left|0\right\rangle_R\right],
\end{equation}
after the normalization constant $\mathcal{N}$ is determined.

Normalized orthogonal left and right moving states are given by
\begin{equation}
\begin{split}
&\left|0\right\rangle_L, \\
&\left|0\right\rangle_R, \\
&\left|\psi^L_a\right\rangle _L=\mathcal{N}^L \psi^{L\dagger}_a\left|0\right\rangle_L, \\
&\left|\psi^R_a\right\rangle _R=\mathcal{N}^R\psi^{R\dagger}_a\left|0\right\rangle_R, \\
\end{split}
\end{equation}
where $(\mathcal{N}^L)^2$ and $(\mathcal{N}^{R})^2$ are given by 
\begin{equation}
\begin{split}
&\left(\mathcal{N}^{L}\right)^{2}= \frac{1}{1+\frac{\left(\gamma^t\gamma^1\right)_{aa}}{2}}, \\
&\left(\mathcal{N}^{R}\right)^2= \frac{1}{1-\frac{\left(\gamma^t\gamma^1\right)_{aa}}{2}}.\\
\end{split}
\end{equation}

A reduced density matrix $\rho_A$ is defined by $\tr_L \rho$. 
It is given by 
\begin{equation}\label{qr}
\rho_A=\frac{1}{4}\left[\left(2+\left(\gamma^t\gamma^1\right)_{aa}\right)\left|0\right\rangle_R\left\langle0\right|_R+\left(2-\left(\gamma^t\gamma^1\right)_{aa}\right)\left|\psi^R_a \right\rangle_R\left\langle\psi^R_a \right|_R \right]
\end{equation}
which agrees with the one obtained by the replica trick. When $\left(\gamma^t\gamma^1\right)_{aa}$ vanishes, $\Delta S^{(n)}_A$ 
are given by $\log{2}$. In the $n \rightarrow 1$ limit, it is given by entanglement entropy for maximally entangled state (EPR state).
If we choose gamma matrices as in (\ref{gtg1}), the anti-particles created by $\psi_1$ is not able to propagate in the right direction parallel to  $x^1$ axis. Therefore the probability with which it is finally included in $B$ is larger than the one with which it is included in $A$ at the late time. The reduced density matrix in (\ref{qr}) agrees with our intuitive expectation.
\subsubsection{$\rho_A$ for $\bar{\psi}\psi$}

A locally excited states is given by 
\begin{equation}\label{bp}
\left |\Psi \right \rangle = \mathcal{N} \bar{\psi}\psi\left|0\right\rangle,
\end{equation}
which is invariant under the $SL(2, {\bf C})$ transformation.

By using an unitary transformation, $\gamma^t\gamma^1$ can be diagonalized. After diagonalize it, the reduced density matrix for this state can be computed easily.
Here it is given by that in (\ref{gtg1})
  
Under the decomposition in (\ref{DQ}), the state in (\ref{bp}) is given by 
\begin{equation}
\left|\Psi\right\rangle =\frac{1}{4}\sum_a\left[\bar{\psi}^{L\dagger}_a \psi^{L\dagger}_a +\bar{\psi}^{L\dagger}_a \psi^{R\dagger}_a+\bar{\psi}^{R\dagger}_a \psi^{L\dagger}_a+\bar{\psi}^{R\dagger}_a \psi^{R\dagger}_a \right] \left| 0 \right \rangle_L \otimes \left|0\right\rangle_R,
\end{equation}
after the normalization constant $\mathcal{N}$ is determined.

Normalized orthogonal states are given by
\begin{equation}
\begin{split}
&\left|0\right\rangle_L, \\
&\left|0\right\rangle_R, \\
&\left|\bar{\psi}^L\psi^L\right\rangle_L=\mathcal{N}^{LL}\sum_a\bar{\psi}_a^{L\dagger}\psi_a^{L\dagger}\left|0\right\rangle_L, \\
&\left|\bar{\psi}^R\psi^R\right\rangle_R=\mathcal{N}^{RR}\sum_a\bar{\psi}_a^{R\dagger}\psi_a^{R\dagger}\left|0\right\rangle_R, \\
&\left|\psi^L_a\right\rangle =\tilde{\mathcal{N}}^L_a\psi^{L\dagger}_a\left|0\right\rangle_L, \\
&\left|\psi^R_a\right\rangle =\tilde{\mathcal{N}}^R_a\psi^{R\dagger}_a\left|0\right\rangle_R, \\
&\left|\bar{\psi}^L_a\right\rangle =\bar{\mathcal{N}}^L_a\psi^{L\dagger}_a\left|0\right\rangle_L, \\
&\left|\bar{\psi}^R_a\right\rangle =\bar{\mathcal{N}}^R_a\psi^{R\dagger}_a\left|0\right\rangle_R, \\
\end{split}
\end{equation}
where normalization factors are given by
\begin{equation}
\begin{split}
&\left(\mathcal{N}^{LL}\right)^2=\left(\mathcal{N}^{RR}\right)^2=\frac{1}{3},\\
&\left(\tilde{\mathcal{N}}^{R}_1\right)^2=\left(\tilde{\mathcal{N}}^{R}_3\right)^2=\left(\tilde{\mathcal{N}}^{L}_2\right)^2=\left(\tilde{\mathcal{N}}^{L}_4\right)^2=
\left(\bar{\mathcal{N}}^{R}_2\right)^2=\left(\bar{\mathcal{N}}^{R}_4\right)^2=\left(\bar{\mathcal{N}}^{L}_1\right)^2=\left(\bar{\mathcal{N}}^{L}_3\right)^2=2, \\
&\left(\tilde{\mathcal{N}}^{R}_2\right)^2=\left(\tilde{\mathcal{N}}^{R}_4\right)^2=\left(\tilde{\mathcal{N}}^{L}_1\right)^2=\left(\tilde{\mathcal{N}}^{L}_3\right)^2=
\left(\bar{\mathcal{N}}^{R}_1\right)^2=\left(\bar{\mathcal{N}}^{R}_3\right)^2=\left(\bar{\mathcal{N}}^{L}_2\right)^2=\left(\bar{\mathcal{N}}^{L}_4\right)^2=\frac{2}{3}.\\
\end{split}
\end{equation}

A reduce density matrix is given by
\begin{equation}
\begin{split}
\rho_A &= \tr_L \rho = \\
&=\frac{12}{64}\left|0\right\rangle_R\left\langle0\right|_R+\sum_{i}\frac{\mathcal{M}_i}{64}\left|\psi^R_i\right\rangle_R\left\langle \psi^R_i\right|_R+\sum_{i}\frac{\bar{\mathcal{M}}_i}{64}\left|\bar{\psi}^R_i\right\rangle_R\left\langle\bar{ \psi}^R_i\right|_R+\frac{12}{64}\left|\bar{\psi}^R\psi^R\right\rangle_R\left\langle\bar{\psi}^R\psi^R \right|_R,
\end{split}
\end{equation}
where $\mathcal{M}_i$ and $\bar{\mathcal{M}}_i$ are given by
\begin{equation}
\begin{split}
&\mathcal{M}_1=\mathcal{M}_3=\bar{\mathcal{M}}_2=\bar{\mathcal{M}}_4=1, \\
&\mathcal{M}_2=\mathcal{M}_4=\bar{\mathcal{M}}_1=\bar{\mathcal{M}}_3=9.
\end{split}
\end{equation}
This reduced density matrix agrees with the one obtained by the replica trick.

\subsubsection{$\rho_A$ for $\psi^{\dagger}\psi$}
A locally excited states is given by 
\begin{equation}\label{dp}
\left |\Psi \right \rangle = \mathcal{N} \psi^{\dagger}\psi \left|0\right\rangle,
\end{equation}
which is variant under the $SL(2, {\bf C})$ transformation.
However $\Delta S^{(n)}_A $ for (\ref{dp}) is invariant in the replica trick. Therefore we diagonalize $\gamma^t\gamma^1$ as in (\ref{gtg1}). After diagonalizing it, $ \rho _A$ is given by 
\begin{equation}
\begin{split}
\rho_A = \tr_R\rho = \frac{20}{64}\left| \phi^R \psi^R \right \rangle_R \left \langle \phi^R \psi^R \right|_R +\frac{20}{64}\left| 0 \right \rangle_R \left \langle 0 \right|_R
+\frac{3}{64}\sum_i \left|\phi^R_i\right\rangle_R\left\langle \phi^R_i \right|_R 
+\frac{3}{64}\sum_i \left|\psi^R_i\right\rangle_R\left\langle \psi^R_i \right|_R.
\end{split}
\end{equation}

\section{Conclusions and Discussions}
In this paper we have studied the dynamics of quantum entanglement. We considered the $4$ dimensional free massless fermionic field theory. Locally excited states are generated by being acted on the ground state by various local operators. We defined the excesses of (R$\acute{e}$nyi) entanglement entropies $\Delta S^{(n)}_A$ by subtracting (R$\acute{e}$nyi) entanglement entropies for the ground state from those for locally excited states. $\Delta S^{(n)}_A$ are finite quantities. 
We found that $\Delta S^{(n)}_A$ vanish if $t\le l$. If $t>l$, $\Delta S^{(n)}_A$ increase and finally approach constants  as in scalar field theory \cite{m1,m2,m3}.

We found that if the locally excited state is given by $\mathcal{N}\psi_a\left|0\right\rangle, \mathcal{N}\bar{\psi}_a\left|0\right\rangle$, components of its reduced density matrices include $\gamma^t \gamma^1$. Therefore they depend on the direction of spin. 
A component of $\psi$ or $\bar{\psi}$ creates an anti-particle or particle. Their propagation depends on the direction of spin. Therefore components of reduced density matrices for those locally excited states depend on the direction of spin.
We also generated  locally excited states $\mathcal{N}\psi^{\dagger }\psi\left|0\right\rangle$ and $\mathcal{N}\bar{\psi}\psi\left|0\right\rangle$  and investigated the time evolution of $\Delta S^{(n)}_A$ for them. We found that $\Delta S^{(n)}_A$ are invariant under the lorentz transformation although $\mathcal{N}\psi^{\dagger }\psi\left|0\right\rangle$ is not invariant under the lorentz transformation. 

We found that the time evolution and the late time values of $\Delta S^{(n)}_A$ can be interpreted in terms of quasi-particles. However in the fermionic field theory, the number of modes which propagate in the left ($x^1<0$) or the right ($x^1>0$) direction depend on the choice of the spin's direction. Therefore we need to impose exotic anti-commutation relation in (\ref{dac}) on quasi-particles. In the $4$ dimensional free massless fermionic field theory, the results by the replica trick can be interpreted as the quantum entanglement between quasi-particles if the relationship in (\ref{dac}) is applied to them. In the $d(\neq 4)$ dimensional free fermionic field theory, anti-commutation relationship should be changed.

We would like to list a few of future problems:
\begin{itemize}
\item[-]{It is interesting to study charged (R$\acute{e}$nyi) entanglement entropies for locally excited stats and holographic field theory.}
\item[-]{It is important to study how we should deform the relationship in (\ref{dac}) in $d(\neq 4)$ dimensional free massless fermionic field theory. }
\item[-] {It is interesting to study the time evolution of $\Delta S^{(n)}_A$ in non-relativistic field theory.}
\item[-] {It would be expected that gamma matrices included in reduced density matrices is related to  the modular Hamiltonian. It is interesting to clarify the relationship between the spin dependence of the reduced matrices and the entanglement Hamiltonian.}
\end{itemize} 
\section*{Acknowledgement}
MN thanks Tadashi Takayanagi and Masaharu Tanahashi for useful discussions and advices. We thank Mitsutoshi Fujita for collaboration in the early stage of this work. 
We also thank Tatsuma Nishioka, Daniel Grumiller, Max Riegler and Pawel Caputa for useful discussions.



\begin{thebibliography}{n}
\bibitem{Holzhey} C. Holzhey, F. Larsen, and F. Wilczek,
``Geometric and Renormalized Entropy in Conformal Field Theory,"
Nucl. Phys. B {\bf 424}, 443 (1994) [hep-th/9403108]

\bibitem{Vidal}
G. Vidal, J. I. Latorre, E. Rico, and A. Kitaev,
``Entanglement in quantum critical phenomena,"
Phys. Rev. Lett. {\bf 90}, 227902 (2003) [quant-ph/0211074]\\
J. I. Latorre, E. Rico, and G. Vidal,
``Ground state entanglement in quantum spin chains,"
Quant. Inf. and Comp. {\bf 4}, 048 (2004) [quant-ph/0304098]

\bibitem{cc-04}
P.~Calabrese and J.~Cardy,
``Entanglement entropy and quantum field theory,"
J. Stat. Mech. P06002 (2004) [hep-th/0405152]

\bibitem{Cal-Lev}
P.~Calabrese, and A.~Lefevre
``Entanglement spectrum in one-dimensional systems,"
Phys. Rev A {\bf 78}, 032329 (2008)
[arXiv:0806.3059 [cond-mat.str-el]].


\bibitem{Kitaev:2005dm} 
  A.~Kitaev and J.~Preskill,
  ``Topological entanglement entropy,''
  Phys.\ Rev.\ Lett.\  {\bf 96}, 110404 (2006)
  [hep-th/0510092].

\bibitem{LW} 
M.~Levin, X.-G.~Wen
``Detecting topological order in a ground state wave function"
Phys. Rev. Lett., 96, 110405 (2006)
arXiv:cond-mat/0510613 [cond-mat.str-el]


\bibitem{RT}
  S.~Ryu and T.~Takayanagi,
  ``Holographic derivation of entanglement entropy from AdS/CFT,''
  Phys.\ Rev.\ Lett.\  {\bf 96} (2006) 181602 [hep-th/0603001]; S.~Ryu and T.~Takayanagi,
 ``Aspects of holographic entanglement entropy,''
  JHEP {\bf 0608} (2006) 045
[hep-th/0605073];
 V.~E.~Hubeny, M.~Rangamani and T.~Takayanagi, ``A Covariant
holographic entanglement entropy proposal,'' JHEP {\bf 0707} (2007)
062  [arXiv:0705.0016 [hep-th]];
T.~Nishioka, S.~Ryu and T.~Takayanagi,
 ``Holographic Entanglement Entropy: An Overview,''
  J.\ Phys.\ A  {\bf 42} (2009) 504008
[arXiv:0905.0932 [hep-th]];
 T.~Takayanagi,
  ``Entanglement Entropy from a Holographic Viewpoint,''
  Class.\ Quant.\ Grav.\  {\bf 29} (2012) 153001  [arXiv:1204.2450 [gr-qc]].


\bibitem{VanRaamsdonk:2010pw} 
  M.~Van Raamsdonk,
  ``Building up spacetime with quantum entanglement,''
  Gen.\ Rel.\ Grav.\  {\bf 42}, 2323 (2010)
  [Int.\ J.\ Mod.\ Phys.\ D {\bf 19}, 2429 (2010)]
  [arXiv:1005.3035 [hep-th]].

\bibitem{arealaw}
M. B. Hastings
``An Area Law for One Dimensional Quantum Systems,''
JSTAT, P08024 (2007)
arXiv:0705.2024 [quant-ph]

\bibitem{econtour}
Yangang Chen, Guifre Vidal
 ``Entanglement contour,''
J. Stat. Mech. (2014) P10011
arXiv:1406.1471 [cond-mat.str-el]

\bibitem{Nozaki:2013wia} 
  M.~Nozaki, T.~Numasawa and T.~Takayanagi,
  ``Holographic Local Quenches and Entanglement Density,''
  JHEP {\bf 1305}, 080 (2013)
  [arXiv:1302.5703 [hep-th]].

\bibitem{out of eq rev}
J. Eisert, M. Friesdorf, C. Gogolin
 ``Quantum many-body systems out of equilibrium,''
Nature Physics 11, 124 (2015)


\bibitem{entanglement-cold}
A. Lamacraft and J. Moore, Ultracold Bosonic and Fermionic Gases, Volume 5 (Contemporary Concepts of Condensed Matter Science) (Elsevier, Oxford, UK, 2012).


\bibitem{cc-global-quench}
P.~Calabrese and J.~L.~Cardy,
  ``Evolution of Entanglement Entropy in One-Dimensional Systems,''
  J.\ Stat.\ Mech.\  {\bf 04} (2005) P04010, cond-mat/0503393.
\bibitem{cc-local-quench}
 P.~Calabrese and J.~L.~Cardy,
  ``Entanglement and correlation functions following a local quench: a conformal field theory approach,''
  J.Phys.A42:504005,2009, [arXiv:0905.4013 [cond-mat.stat-mech]];
 V.~ Eisler, F.~ Igloi, and I.~ Peschel, 
``Evolution of entanglement after a local quench,''
J. Stat. Mech. P02011 (2009), [cond-mat/0703379 [cond-mat.stat-mech]].

 \bibitem{Alc} 
  F.~C.~Alcaraz, M.~I.~Berganza and G.~Sierra,
  ``Entanglement of low-energy excitations in Conformal Field Theory,''
  Phys.\ Rev.\ Lett.\  {\bf 106}, 201601 (2011)
  [arXiv:1101.2881 [cond-mat.stat-mech]].

\bibitem{m1} 
  M.~Nozaki, T.~Numasawa and T.~Takayanagi,
  ``Quantum Entanglement of Local Operators in Conformal Field Theories,''
  Phys.\ Rev.\ Lett.\  {\bf 112}, 111602 (2014)
  [arXiv:1401.0539 [hep-th]].
\bibitem{m2} 
M.~Nozaki,
  ``Notes on Quantum Entanglement of Local Operators,''
  JHEP {\bf 1410}, 147 (2014)
  [arXiv:1405.5875 [hep-th]].
\bibitem{m3}
  P.~Caputa, M.~Nozaki and T.~Takayanagi,
  ``Entanglement of local operators in large-N conformal field theories,''
  PTEP {\bf 2014}, no. 9, 093B06 (2014)
  [arXiv:1405.5946 [hep-th]].
  \bibitem{He} 
  S.~He, T.~Numasawa, T.~Takayanagi and K.~Watanabe,
  ``Quantum dimension as entanglement entropy in two dimensional conformal field theories,''
  Phys.\ Rev.\ D {\bf 90}, no. 4, 041701 (2014)
  [arXiv:1403.0702 [hep-th]].
  
  \bibitem{He2} 
  B.~Chen, W.~Z.~Guo, S.~He and J.~q.~Wu,
  ``Entanglement Entropy for Descendent Local Operators in 2D CFTs,''
  arXiv:1507.01157 [hep-th].
  
  \bibitem{Cap2} 
  P.~Caputa and A.~Veliz-Osorio,
  ``Entanglement constant for conformal families,''
  arXiv:1507.00582 [hep-th].
  
  \bibitem{Cap} 
  P.~Caputa, J.~Simon, A.~stikonas and T.~Takayanagi,
  ``Quantum Entanglement of Localized Excited States at Finite Temperature,''
  JHEP {\bf 1501}, 102 (2015)
  [arXiv:1410.2287 [hep-th]].
  \bibitem{Hart} 
  C.~T.~Asplund, A.~Bernamonti, F.~Galli and T.~Hartman,
  ``Holographic Entanglement Entropy from 2d CFT: Heavy States and Local Quenches,''
  JHEP {\bf 1502}, 171 (2015)
  [arXiv:1410.1392 [hep-th]].

\bibitem{cree} 
  A.~Belin, L.~Y.~Hung, A.~Maloney, S.~Matsuura, R.~C.~Myers and T.~Sierens,
  ``Holographic Charged Renyi Entropies,''
  JHEP {\bf 1312}, 059 (2013)
  [arXiv:1310.4180 [hep-th]].


\bibitem{thermo-ent}
J. Bhattacharya, M. Nozaki, T. Takayanagi and T. Uga- jin, Phys. Rev. Lett. 110, no. 9, 091602 (2013) [arXiv:1212.1164 ]
D. D. Blanco, H. Casini, L. -Y. Hung and R. C. My- ers, JHEP 1308, 060 (2013) [arXiv:1305.3182 [hep-th]]; G. Wong, I. Klich, L. A. Pando Zayas and D. Vaman,
arXiv:1305.3291 [hep-th]
F. C. Alcaraz, M. I. Berganza, G. Sierra, Phys. Rev. Lett.
106 (2011) 201601 [arXiv:1101.2881 [cond-mat]]

\bibitem{Dow} 
  J.~S.~Dowker,
  ``Quantum Field Theory on a Cone,''
  J.\ Phys.\ A {\bf 10}, 115 (1977);M.~E.~X.~Guimaraes and B.~Linet,
  ``Scalar Green's functions in an Euclidean space with a conical-type line singularity,''
  Commun.\ Math.\ Phys.\  {\bf 165}, 297 (1994).
\bibitem{sac2}
M.~A.~Metlitski, C.~A.~Fuertes, and S.~Sachdev,``Entanglement Entropy in the $O(N)$ model,''
 Phys.\ Rev.\ B {\bf
80},115122 (2009) [arXiv:0904.4477 [cond-mat.stat-mech]].
\bibitem{Line} 
  B.~Linet,
  ``Euclidean spinor Green's functions in the space-time of a straight cosmic string,''
  J.\ Math.\ Phys.\  {\bf 36}, 3694 (1995)
  [gr-qc/9412050].
\bibitem{Nozaki:2014hna} 
  M.~Nozaki, T.~Numasawa and T.~Takayanagi,
  ``Quantum Entanglement of Local Operators in Conformal Field Theories,''
  Phys.\ Rev.\ Lett.\  {\bf 112}, 111602 (2014)
  [arXiv:1401.0539 [hep-th]].






\bibitem{chm}
H.~Casini, M.~Huerta and R.~C.~Myers,
  ``Towards a derivation of holographic entanglement entropy,''
  JHEP {\bf 1105}, 036 (2011)
  [arXiv:1102.0440 [hep-th]].
  \bibitem{BHMMMS}
A.~Belin, L.~Y.~Hung, A.~Maloney, S.~Matsuura, R.~C.~Myers and T.~Sierens,
``Holographic Charged R\'enyi Entropies,''
JHEP {\bf1312} (2013) 059,
[arXiv:1310.4180 [hep-th]].


\end{thebibliography}
\end{document}